\documentclass[a4paper,11pt]{article}
\usepackage[right]{lineno}
\usepackage{amsmath,amssymb,color}
\usepackage[dvipdfmx]{graphicx}
\usepackage{hyperref}
    \hypersetup{
        colorlinks   = true,
        citecolor    = blue
    }

\usepackage{amsthm,bm,comment}
\usepackage{here}
\usepackage{color}

\newcommand{\red}[1]{\textcolor{red}{#1}}

\usepackage{authblk} 
\usepackage{natbib}

\theoremstyle{plain}
\newtheorem{thm}{Theorem}
\newtheorem{ass}{Assumption}

\newtheorem{lem}{Lemma}
\theoremstyle{definition}
\newtheorem{rem}{Remark}
\newtheorem*{rem*}{Remark}

\newcommand{\argmax}{\mathop{\rm arg max}\limits}

\definecolor{ogata}{rgb}{.7,0,1}

\title{A Mixture Transition Distribution Modeling for Higher-Order Circular Markov Processes}

\author[1]{Hiroaki Ogata\thanks{1-1 Minami-Osawa, Hachioji, Tokyo, 192-0397, Japan. \href{email:hiroakiogata@tmu.ac.jp}{hiroakiogata@tmu.ac.jp}}} 
\author[2]{Takayuki Shiohama\thanks{18 Yamazato-cho, Showa, Nagoya, 466-8673, Japan.}}

\affil[1]{Faculty of Economics and Business Administration, Tokyo Metropolitan University}
\affil[2]{Department of Data Science, Nanzan University}

\date{\empty}

\begin{document}

\maketitle

\abstract{This study considers the stationary higher-order Markov process for circular data by employing the mixture transition distribution modeling. The underlying circular transition distribution is based on Wehrly and Johnson's bivariate joint circular models. The structures of the circular autocorrelation function, together with the circular partial autocorrelation function, are investigated. They are found to be similar to those of the autocorrelation and partial autocorrelation functions of the real-valued autoregressive process when the underlying binding density has zero sine moments. The validity of the model is assessed by applying it to some Monte Carlo simulations and real directional data.\\
\noindent
\textbf{keywords:} circular statistics, Markov process, maximum likelihood estimation, spectral density, stationary process, time-series analysis}

\clearpage

\section{Introduction}
Statistical analysis related to the data that take values on unit vectors is called directional statistics, where the direction is more important than the magnitude. Directional data occur in many areas, including geostatistics, natural sciences, environmental sciences, biology, information science, and genetics. The basic manifold of the high dimensional sphere is the unit circle embedded in the Euclidean space $\mathbb{R}^2$, and the statistical methods for the unit circle are known as circular statistics. For further reviews, we refer to the monographs of \cite{jammalamadaka2001topics}, \cite{MJ09}, \cite{pewsey2013circular}, \cite{ley2017modern} and \cite{ley2020applied}. \cite{pewsey2021recent} should also be referred to.

While many data which take values on the unit circle or sphere are characterized by time-series structure, the statistical analysis for circular time series has not been fully developed yet. There are many time-series data on the circle, such as the wind or wave directions, time records of a certain event, and animal movement trajectories, which show typical periodic patterns. As for the correlation coefficient of the bivariate circular data, \cite{fisher1983correlation, fisher1994time} proposed circular correlation coefficients and considered basic circular time-series models as linked autoregressive and moving average models and wrapping processes for real-valued time series. A comprehensive analysis of circular time-series modeling can be found in \cite{breckling1989analysis}.

A major approach for modeling circular time series is to apply the Markov models to it. \cite{wehrly1980bivariate} proposed joint distribution for the circular random variables. \cite{holzmann2006hidden} applied it to time-series modeling using hidden Markov models. \cite{abe2017circular} showed the circular autocorrelation structure of the circular Markov process proposed by \cite{wehrly1980bivariate}. Another Markov-based circular time series was considered by \cite{kato2010markov}; he considered the M\"obius transformation of the circular random variables to produce the transition densities. His models were extended by \cite{jones2015class}, who considered the circular joint distributions called `circulas' to analyze correlated structures for circular random variables. 

While all these approaches are based on the first-order Markov modeling for circular time series, investigating higher-order Markov models on the circle is of primary importance. To the best of the authors' knowledge, no approaches exist in the literature for higher-order circular Markov models. For the statistical modeling with real-valued time series, \cite{raftery1985model} considered the higher-order Markov modeling using a mixture of transition densities (MTD), and his models have been lately developed to the MTD modeling for a higher-order dependency of the observed time series. See for details, \cite{raftery1994estimation}, \cite{le1996modeling}, and \cite{berchtold2002mixture}. In this study, we develop the higher-order Markov process on the circle using the transition density of \cite{wehrly1980bivariate}'s model together with the MTD approaches. We introduce the maximum likelihood estimation for the model parameters and discuss some information criteria for model selection.

The rest of the paper is organized as follows. Section~\ref{sec:2} introduces our proposed higher-order Markov process on the circle by incorporating MTD modeling approaches. The first- and second-order stationarity of the process is investigated. The circular autocorrelation (CACF) and circular partial autocorrelation function (CPACF) are derived in Section~\ref{sec:3}. Moreover, the spectral density function of the higher-order circular time series is also given. In Section~\ref{sec:4}, the maximum likelihood estimator (MLE) is investigated for the unknown parameter estimation. We provide asymptotic normality of the MLE. Section~\ref{sec:5} illustrates some finite sample performances of MLE via Monte Carlo simulations. Real-data analysis is conducted in Section~\ref{sec:6}. Finally, Section~\ref{sec:7} concludes our paper. All proofs of theorems and lemmas are given in Appendix~\ref{sec:A}.

\section{Mixture Transition Distribution Models on a Circle }\label{sec:2}

We consider the circular strictly stationary process $ \{ \Theta_t \}_{t \in \mathbb{Z}} $. Throughout the paper, we use the notation $\theta_{a:b}=\{\theta_a, \theta_{a-1}, \ldots, \theta_{b+1}, \theta_{b}\}$ for the $\sigma$-field generated by the random variables $\{\Theta_t\}, t \in \{ a,a-1, \ldots, b+1,b\}$ for $a \geq b$. The transition density for the Markov process on the circle proposed by \cite{wehrly1980bivariate} was set as
\begin{align}
f(\theta_t |\theta_{t-1: -\infty}) 
=f(\theta_t | \theta_{t-1})
=2\pi g[2\pi \{F(\theta_t)-qF(\theta_{t-1})\}]f(\theta_t).
\label{eq:general}
\end{align}
Here, $ f $ and $ g $ are densities on the circle, $ F(\theta)=\int_{0}^{\theta} f(\xi) d\xi $ is a distribution function, and $ q \in \{-1, 1\}$ is a non-random factor that determines the sign of $F(\theta_{t-1})$ and the direction of the association of consecutive sequences. $ f $ corresponds to the marginal density function of $ \Theta_t $. If we set the marginal as the circular uniform, $ f(\theta_t)=1/(2\pi) $, the above transition density has a simpler form:
\begin{align}\label{eq:TD1}
f(\theta_t | \theta_{t-1}) = g(\theta_t-q\theta_{t-1}).
\end{align}
The density $g$ is sometimes called a binding density as in \cite{jones2015class}. 

Now, we extend the circular Markov process proposed by \cite{wehrly1980bivariate} to a higher-order Markov process. We employ the mixture transition distribution (MTD) model that originates in \cite{raftery1985model}. We use the term $p$th-order Markov process, which indicates that the conditional distribution is determined by only $p$ adjacent past values. 
Referring to (\ref{eq:TD1}), we define the circular $p$th-order mixture transition distribution model by setting the transition density as
\begin{align}\label{eq:MTD}
f(\theta_t |  \theta_{t-1: -\infty}) 
=f(\theta_t | \theta_{t-1:t-p})
=\sum_{i=1}^{p} a_i g(\theta_t - q_i\theta_{t-i}).
\end{align} 
Here, $g(\cdot)$ is a circular density function, and $ a_i~(i=1,\ldots, p-1)\geq 0$ and $a_p >0$ are non negative and positive constants, respectively. The sum of mixing weights must satisfy $\sum_{i=1}^{p} a_i = 1$, and $ q_i $s are constants that take the values $ -1 $ or $ 1 $. Hereafter, we call model (\ref{eq:MTD}) an MTD-AR($p$) model. Let the binding density $g$ have the following $m$-th trigonometric moment: 
\begin{align*}
\mathbb{E}_g[\exp ( \mathrm{i}m\Theta)] = \rho_m e^{ \mathrm{i} \mu_m}
\end{align*}
where $\mathrm{i}$ is an imaginary unit defined as $\mathrm{i}^2 =-1$; 
$\rho_m 
$ and $\mu_m 
$ are the $m$-th order mean resultant length and mean direction; 
respectively. Now we propose the following assumption. 
\begin{ass}
\label{ass1}
\begin{itemize}
\item[(a)] 
For positive integer $m$, the m-th order mean resultant length of the binding density is positive and less than unity, that is $0< \rho_m < 1$ for $m=\pm 1,\pm 2,\ldots$.
\item[(b)]\label{ass:1b} $\sum_{m=-\infty}^{\infty}\rho_m^2 < \infty$. 
\end{itemize}
\end{ass}

Assumption \ref{ass1}(b) ensures that the binding density $g$ has the form as
\begin{align*}
g(\theta)=
\frac{1}{2\pi}
\sum_{m=-\infty}^{\infty} \rho_m e^{-\mathrm{i} m \theta}.
\end{align*}
This assumption is crucial for the ergodicity of the MTD-AR($p$) process and deriving CACF and CPACF together with the circular spectral density function discussed in Section~\ref{sec:3}. 
\cite{di2012non} and \cite{beran2022long} discussed the case when $\sum_{m=-\infty}^{\infty}\rho_m^2=\infty$, which yields an angular valued long range dependent process. 

When $m=1$ and $q_1=1$, the mean direction $\mu_1$ determines the local trend or direction of the movement in angular around the previous direction $\theta_{t-1}$ with concentration $\rho_1$. For the case with $q_1=-1$, describing the role of the parameter $\mu_1$ is difficult to interpret.

Now we introduce the relationship
\begin{align*}
	\int_{\Pi} 
	\left(
	\begin{array}{c}
	\cos m\theta_j \\
	\sin m\theta_j
	\end{array}
	\right)
	g(\theta_j-q_i \theta_{j-i}) d\theta_j
	=D_m
    Q_i
	\left(
	\begin{array}{c}
	\cos m\theta_{j-i} \\
	\sin m\theta_{j-i}
	\end{array}
	\right),
\end{align*}
where $\Pi=(-\pi, \pi]$ and
\begin{align*}
D_m = 
\rho_m
\begin{pmatrix}
\cos \mu_m & -\sin\mu_m\\
\sin\mu_m & \cos\mu_m 
\end{pmatrix}
~~~\textrm{and}~~~
Q_i= 
\begin{pmatrix}
1 &  0\\
0&  q_i
\end{pmatrix}.
\end{align*}
The matrix $Q_i$ determines the direction of the association between $\Theta_t$ and $\Theta_{t-i}$. See Lemma~1 in the Appendix for derivation. The matrix $D_m$ plays an important role in determining the autocorrelation structures of the Markov processes.

\subsection{First- and second-order stationarity}\label{sec:2.1}

The process is said to be second-order stationary if its first and second moments are all independent of the time $t$. First, we consider the conditions for the process $ \{\bm{U}_t = (\cos\Theta_t \ \sin\Theta_t)^{T}\}_{t \in \mathbb{Z}} $ to be the first-order stationary. Let us denote $ \bm{\mu}_{t}=E(\bm{U}_t)$, and $\bm{m}_t = ( \bm{\mu}_{t}^T,\ldots,  \bm{\mu}_{t-p+1}^T)^T$. Then,
\begin{align*}
\bm{\mu}_{t}
=
E\left[
E\left(
\left.
\bm{U}_{t}
\right|
\Theta_{t-1: t-p}
\right)
\right]
=
E\left[
\sum_{i=1}^{p} a_i D_1 Q_i
\bm{U}_{t-i}
\right] 
=
\sum_{i=1}^{p} a_i D_1 Q_i
\bm{\mu}_{t-i}.
\end{align*} 
For the detailed derivation of this expression, see Lemma \ref{lemma 1} in the Appendix. In matrix form, this is written as
\begin{align} 
\bm{m}_t=
\left(
\begin{array}{l}
\bm{\mu}_{t} \\
\bm{\mu}_{t-1} \\
\multicolumn{1}{c}{\vdots} \\
\bm{\mu}_{t-p+1}
\end{array}
\right)
= 
\left(
\begin{array}{cccc}
a_1D_1 Q_1
& \cdots & a_{p-1}D_1 Q_{p-1}
& a_{p}D_1 Q_p
 \\
I_2 & \cdots & O & O  \\
\vdots & \ddots & \vdots & \vdots  \\
O & \cdots & I_2 & O  \\
\end{array}
\right)   
\left(
\begin{array}{l}
\bm{\mu}_{t-1} \\
\bm{\mu}_{t-2} \\
\multicolumn{1}{c}{\vdots} \\
\bm{\mu}_{t-p}
\end{array}
\right)
 =:
\bm{A}\bm{m}_{t-1},
\label{matrix form}
\end{align}
where $I_2$ denotes the $ 2 \times 2 $ identity matrix. Following the proof of Theorem 1 in \citet{le1996modeling}, we find that if and only if all the absolute value of the eigenvalues of $ 2p \times 2p $ matrix $\bm{A}$ defined by equation (\ref{matrix form}) are less than $ 1 $, then the process $ \{\bm{U}_t\}_{t \in \mathbb{Z}} $ is first-order stationary. The required condition for the first-order stationarity is that all the roots of the characteristic equation
\begin{align*}
 \det( \lambda^{p} I_2  - \lambda^{p-1} a_1 D_1Q_1 
 -  \cdots - a_{p} D_1Q_p ) =0
\end{align*}
lie in the unit circle.

\begin{rem}[MTD-AR(1) case]
The first-order stationary condition is the absolute values of the eigenvalues of $ D_1$, which are both less than 1. Recall that $a_1=1$ if $p=1$ by definition. The characteristic polynomial is
\begin{align*}
&\det(\lambda I_2- D_1Q_1) 
=\det
\left( 
\begin{array}{cc}
\lambda-\rho_1 \cos\mu_1 & q_1\rho_1 \sin\mu_1 \\
-\rho_1 \sin\mu_1 & \lambda-q_1\rho_1 \cos\mu_1
\end{array}
\right) \\
=&\lambda^2-(q_1+1) \rho_1\cos\mu_1  \cdot \lambda + q_1 (\rho_1)^2.
\end{align*} 
The roots of the characteristic equation are $ \rho_1 (\cos\mu_1 \pm i \sin\mu_1)$ for $q_1=1$ and $\pm \rho_1$ for $q_1=-1$. For both cases with $q_1 = \pm 1$, the absolute values of the roots are less than one unless the distribution $g$ degenerates. Assumption 1 ensures that the process is first-order stationary. 
\end{rem}
\begin{rem}[MTD-AR(2) case]
For the MTD-AR(2) model, the characteristic polynomial becomes
\begin{align}
\nonumber
&\det(\lambda^2 I_2- \lambda a_1D_1Q_1 -a_2 D_1Q_2)
\nonumber
=\lambda^4 
-(q_1+1)a_1\rho_1\cos \mu_1 \lambda^3\\
+&(q_1a_1^2\rho_1^2 -(q_2+1)a_2 \rho_1\cos \mu_1 )\lambda^2
+ 2a_1a_2  \rho_1^2 \lambda
+ a_2^2 \rho_1^2(q_1\cos^2\mu_1+q_2\sin^2\mu_1). 
\label{eq:ar2poly}
\end{align}
Hence, the roots of (\ref{eq:ar2poly}) can be uniquely determined when the density function $g$ is specified, and the absolute values of the resulting roots are less than unity, which indicates the MTD-AR(2) model is always first-order stationary.  
\end{rem}

Next, we consider second-order stationarity of the process $\{\bm{U}_t \}_{t \in \mathbb{Z}} $. Let us denote $ V_{t}=E(  \bm{U}_t \bm{U}_t^{T} ) $.
Then,
\begin{align}
\label{eq:Vt}
V_{t}
=&
E\left[
E\left\{
\left.
\left(
\begin{array}{cc}
\cos^2 \Theta_t & \cos \Theta_t \sin \Theta_t \\ 
\sin \Theta_t \cos \Theta_t & \sin^2 \Theta_t  
\end{array}
\right) 
\right|
\Theta_{t-1: t-p}
\right\}
\right] \nonumber \\
=&
E\left[
E\left\{
\left.
\frac{1}{2}I_2
+\frac{1}{2}
\left(
\begin{array}{cc}
\cos 2\Theta_t &  \sin 2\Theta_t \\ 
\sin 2\Theta_t & -\cos 2\Theta_t  
\end{array}
\right) 
\right|
\Theta_{t-1:t-p}
\right\}
\right] \nonumber \\
=&
E\left[
\frac{1}{2}I_2
+\frac{1}{2}
\sum_{i=1}^{p} a_i D_2Q_i
\left(
\begin{array}{cc}
\cos 2\Theta_{t-i} &  \sin 2\Theta_{t-i}  \\ 
\sin 2\Theta_{t-i} & -\cos 2\Theta_{t-i}  
\end{array}
\right) Q_i
\right] \nonumber \\
=& 
\frac{1}{2}
\left(
I_2-D_2
\right)+
D_2
\sum_{i=1}^{p} a_i Q_i V_{t-i} Q_i.
\end{align} 
 For the above derivation, we use Lemma~\ref{lemma 2} in the Appendix to calculate the conditional expectation. We define the following $2p \times 2p$ block matrices:
\begin{align*}
\mathcal{V}_t
=&
\left(
\begin{array}{cccc}
V_t &  &  & \\
 & V_{t-1} &  &   \\
 & & \ddots &   \\
 & & & V_{t-(p-1)}  \\
\end{array}
\right), ~
\mathcal{D}
=
\left(
\begin{array}{cccc}
 D_2 & O
& \cdots &  O
 \\
O & I_2 & \cdots & O  \\
\vdots & \vdots & \ddots & \vdots  \\
O & O & \cdots & I_2  \\
\end{array}
\right), \\
\mathcal{Q}
=&
\left(
\begin{array}{cccc}
\sqrt{a_1} Q_1
& \cdots & \sqrt{a_{p-1}} Q_{p-1}
& \sqrt{a_{p}} Q_p
 \\
I_2 & \cdots & O & O  \\
\vdots & \ddots & \vdots & \vdots  \\
O & \cdots & I_2 & O  \\
\end{array}
\right),
\end{align*} 
and $2p \times 2$ block matrix
\begin{align*}
\mathcal{C}
=
\left(
\begin{array}{cccc}
\frac{1}{2}(I_2-D_2^T) &
O &
\cdots &
O  
\end{array}
\right)^T. 
\end{align*} 
 Then, (\ref{eq:Vt}) is written as
 \begin{align}
 \label{eq:Wt}
 \mathcal{V}_t=\mathcal{D}\mathcal{Q} \mathcal{V}_{t-1} \mathcal{Q}^{T}
 +\mathcal{C}.
 \end{align}
Using vec operator and Kronecker product, (\ref{eq:Wt}) is written as
 \begin{align*}
\mathrm{vec}(\mathcal{V}_t) = (\mathcal{Q} \otimes \mathcal{D}\mathcal{Q}) \mathrm{vec}(\mathcal{V}_{t-1}) + \mathrm{vec}(\mathcal{C}).
\end{align*}
The second-order stationarity of $\bm{U}_{t}$ requires that all absolute values of the eigenvalues of the $4p^2 \times 4p^2$ matrix $\mathcal{Q} \otimes \mathcal{D}\mathcal{Q}$ are less than 1. The eigenvalues of $\mathcal{Q} \otimes \mathcal{D}\mathcal{Q}$ are of the form $\lambda_i\nu_j$ where $\lambda_i$ and $\nu_j$ are eigenvalues of $\mathcal{Q}$ and $\mathcal{D}\mathcal{Q}$. If $|\lambda_i|<1$ and $|\nu_j|<1$ for all $i$ and $j$, all the eigenvalues of $\mathcal{Q} \otimes \mathcal{D}\mathcal{Q}$ are inside the unit circle. Hence, the required condition for the second-order stationarity is similar to that of the first-order condition and is the product of all the combinations of the absolute values of the root of the equations
\begin{align*}
 \det( \lambda^{p} I_2  - \lambda^{p-1}a_1^{1/2}Q_1 -  \cdots - a_{p}^{1/2}Q_p ) =0
\end{align*}
and
\begin{align*}
 \det( \lambda^{p} I_2  - \lambda^{p-1}a_1^{1/2}D_2Q_1 -  \cdots - a_{p}^{1/2}D_2Q_p ) =0
\end{align*}
are less than unity.

\begin{rem}[MTD-AR(1) case]
From $\det \left( \lambda I_2-Q_1 \right)=0$, we obtain eigenvalues as $1$ and $q_1$, which lie on the unit circle. Hence, for the second-order stationarity, we need the eigenvalues of $\det\left(\lambda I_2 - D_2Q_1  \right)$ to be less than 1. The characteristic polynomial is
	\begin{align*}
	&\det\left(\lambda I_2 - D_2Q_1  \right) 
	=\det
	\left( 
	\begin{array}{cc}
	\lambda-\rho_2 \cos\mu_2  & q_1\rho_2  \sin\mu_2  \\
	-\rho_2 \sin\mu_2  & \lambda-q_1\rho_2 \cos\mu_2 
	\end{array}
	\right) \\
	=&\lambda^2-
	\rho_2\cos \mu_2(1+ q_1)\lambda
+	
q_1 \rho_2^2.
	\end{align*}
As the roots of the characteristic equation are $ \rho_2 (\cos\mu_2 \pm \textrm{i} \sin\mu_2) $ if $q_1=1$ and $ \pm\rho_2$ if $q_1=-1$. For both cases, the roots are less than 1, and the process is always second-order stationary.
\end{rem}

There are several examples of non-stationary process of the MTD-AR($p$) related models. First, if $\rho_1=1$, that is, the binding density $g$ degenerates to a certain direction $\mu_g$, and the process becomes deterministic. For example, let us consider the case of $p=1$ and $q_1=1$ and set the initial value $\Theta_0=\theta_0$. Then, $\theta_t=\theta_0+\mu_g t$ $(t=1,2,\ldots)$. This process is not first-order stationary, as the mean of the process depends on time $t$. When $\mu_g=0$, the process takes the same value as the initial direction $\theta_0$. This process is stationary but not uniformly ergodic, as all values depend on the initial value. 
The second example is time-varying concentration parameter models, that is, the binding density has time-varying concentration $\rho_1(\theta_{t-1})$ that is, 
 concentration depends on the past value $\theta_{t-1}$.
 
So far, we have discussed the first- and second-order stationarity of the bivariate trigonometric process $\bm{U}_t = (\cos \Theta_t, \sin \Theta_t)^T$ induced from the MTD-AR($p)$ process defined by $\{\Theta_t\}_{t \in \mathbb{Z}}$. The following remark refers to the strictly stationary results of the angular valued process $\{\Theta_t\}_{t \in \mathbb{Z}}$.

\begin{rem}
Let $\pi(\theta_1,\ldots \theta_k)$ be a $k$-dimensional probability distribution of a random vectors $\theta_1,\ldots, \theta_k$.
From the definition of the transition density given by (\ref{eq:MTD}), the  joint distribution $\pi(\theta_t, \theta_{t-1}, \ldots, \theta_{t-p})$ becomes
\begin{align*}
&\pi(\theta_t, \theta_{t-1}, \ldots, \theta_{t-p})
=
\left(
\sum_{j=1}^{p} 
a_j g(\theta_t-\theta_{t-j}) 
\right)
\pi(\theta_{t-1}, \ldots, \theta_{t-p}) \\
=&
\sum_{j=1}^{p} 
a_j g(\theta_t-\theta_{t-j}) 
\pi(\theta_{t-j}) 
=
\frac{1}{2\pi}
\sum_{j=1}^{p} 
a_j g(\theta_t-\theta_{t-j}).
\end{align*}
This implies the stationary distribution of $\theta_t$ becomes a circular uniform distribution, as
\begin{align*}
 &\pi(\theta_t)
=
\int_{-\pi}^{\pi} 
\cdots
\int_{-\pi}^{\pi} 
\left\{
\frac{1}{2\pi}
\sum_{j=1}^{p} 
a_j g(\theta_t-\theta_{t-j})
\right\}
d\theta_{t-1}
\cdots
d\theta_{t-p} \\
=&
\frac{1}{2\pi}
\sum_{j=1}^{p} 
\left\{
\int_{-\pi}^{\pi} 
a_j g(\theta_t-\theta_{t-j})
d\theta_{t-j}
\right\} 
=
\frac{1}{2\pi}
\sum_{j=1}^{p} 
a_j=\frac{1}{2\pi}.
\end{align*}
Hence, the MTD-AR($p$) process becomes strictly stationary and ergodic.
\end{rem}

\section{Time-Series Structures for Circular Data}\label{sec:3}

We first review the measure of association between two circular random variables $ \Theta $ and $ \Phi $. Most of the existing measures of association involve random vectors $ \bm{U}=(\cos\Theta, \sin\Theta)^{T} $ and $ \bm{V}=(\cos\Phi, \sin\Phi)^{T} $ in the plain. We introduce the following notations:
\begin{align*}
&S_{\Theta\Theta}
=E( \bm{U}\bm{U}^{T} )
=E
\left(
\begin{array}{cc}
\cos^2\Theta & \cos\Theta\sin\Theta \\
\sin\Theta\cos\Theta & \sin^2\Theta 
\end{array}
\right), \\
&S_{\Theta\Phi}
=E(\bm{U}\bm{V}^{T})
=E
\left(
\begin{array}{cc}
\cos\Theta\cos\Phi & \cos\Theta\sin\Phi \\
\sin\Theta\cos\Phi & \sin\Theta\sin\Phi 
\end{array}
\right), \\
&S_{\Phi\Theta}
=E(\bm{V}\bm{U}^{T})
=E
\left(
\begin{array}{cc}
\cos\Phi\cos\Theta & \cos\Phi\sin\Theta \\
\sin\Phi\cos\Theta & \sin\Phi\sin\Theta 
\end{array}
\right), \\
&S_{\Phi\Phi}
=E(\bm{V}\bm{V}^{T})
=E
\left(
\begin{array}{cc}
\cos^2\Phi & \cos\Phi\sin\Phi \\
\sin\Phi\cos\Phi & \sin^2\Phi 
\end{array}
\right). 
\end{align*} 

\subsection{Circular autocorrelation functions}\label{sec:3.1}

The circular correlation coefficient proposed by \citet{fisher1983correlation} is written as
\begin{align*}
\rho_{\mathrm{FL}}(\Theta, \Phi)
=\frac{\det(S_{\Theta\Phi})}{\{\det(S_{\Theta\Theta})\det(S_{\Phi\Phi})\}^{1/2}}.
\end{align*}
Because of the stationarity, the quantities $ \bm{\mu}=E(\bm{U}_t) $ and $ \Gamma_k=E(\bm{U}_{t+k}\bm{U}_t^{T}) $ do not depend on time $ t $.
Then, as proposed in \citet{holzmann2006hidden}, the CACF) at lag $ k $ is defined by
\begin{align}\label{CACF}
r^{(C)}_{k}
=\rho_{\mathrm{FL}}(\Theta_{t+k}, \Theta_t)
=\frac{\det E( \bm{U}_k\bm{U}_0^{T})}{\det E( \bm{U}_0\bm{U}_0^{T} )}.
\end{align}
Now, we give the CACF of (\ref{eq:MTD}). Note that $ \Gamma_{-k}=\Gamma_k^{T} $ and 
\begin{align*}
\Gamma_0
&=E\left[\left(
\begin{array}{cc}
\cos^2 \Theta_0  & \cos \Theta_0 \sin \Theta_0 \\
\sin \Theta_0 \cos \Theta_0 & \sin^2 \Theta_0
\end{array}
\right)\right]
=\frac{1}{2}
E\left[\left(
\begin{array}{cc}
\cos 2\Theta_0+1  & \sin 2\Theta_0 \\
\sin 2\Theta_0 & -\cos 2\Theta_0+1
\end{array}
\right)\right] \\
&=\frac{1}{2}\left(
\begin{array}{cc}
1+\alpha_2^{(f)}  & \beta_2^{(f)} \\
\beta_2^{(f)} & 1-\alpha_2^{(f)}
\end{array}
\right)
= 
\frac{1}{2} I_2,
\end{align*}
where $ \alpha_2^{(f)} $ and $ \beta_2^{(f)} $ are the second trigonometric moment of the marginal distribution $f$ in (\ref{eq:general}), that is, $E_f[e^{\textrm{i} 2\Theta}]= \alpha_2^{(f)} + \textrm{i} \beta_2^{(f)}$. As we assume $f$ to be a circular uniform distribution, we have $\alpha_2^{(f)}=0$ and $ \beta_2^{(f)} =0$. Then, $ \Gamma_k $ can be obtained recursively as follows: 
\begin{align}
\notag
\Gamma_k
=&E\left(
\bm{U}_t 
\bm{U}_{t-k}^{T}
\right) 
=E\left[
E\left(
\bm{U}_t
\bm{U}_{t-k}^{T}
|
\Theta_{t-1:t-p}
\right)
\right]\\
\notag
=&E\left[
E\left(
\bm{U}_t|
\Theta_{t-1:t-p}
\right)
\bm{U}_{t-k}^{T}
\right]
=E\left[
\sum_{i=1}^{p}
a_i D_{1}Q_i \bm{U}_{t-i} \bm{U}_{t-k}^{T}
\right]\\
=& \sum_{i=1}^{p} a_i D_{1}Q_i \Gamma_{k-i}.
\label{eq:recursive_relation}
\end{align}
The following theorem gives the CACF for the $p$-th order circular Markov processes. The proof is omitted, as it is obvious from the above discussion.

\begin{thm}\label{th1}
Assume that Assumption 1 holds, the lag $k$ CACF of the process $ \{\Theta_t \}_{t \in \mathbb{Z}} $ given by (\ref{CACF}) becomes
\begin{align}
r^{(C)}_k
= 
\frac{\det \Gamma_k}{\det \Gamma_0},
\label{eq:CACF}
\end{align}
where $\Gamma_k$ is obtained recursively using (\ref{eq:recursive_relation}) with $\Gamma_0= \frac{1}{2}I_2$. 
\end{thm}

The sample CACF at lag $k$ is defined by
\begin{align*}
\hat{r}^{(C)}_k
= 
\frac{ \det \widehat{\Gamma}_k}{\det \widehat{\Gamma}_0},
\end{align*}
where
\begin{align}
\widehat{\Gamma}_k=
\frac{1}{n-k} \sum_{t=k+1}^n
\bm{U}_{t}\bm{U}_{t-k}^T.
\label{eq:samplecovarinace2}
\end{align}

The obtained CACF is a natural extension of the result given by \citet{holzmann2006hidden}, where they considered the von Mises distribution for $g$ and the first-order Markov process with order $p=1$. \citet{abe2017circular} studied the arbitrary binding density function for $g$ and obtained similar CACF as that obtained by \cite{holzmann2006hidden}. The result given in Theorem~\ref{th1} is the CACF for general $p$-th order circular Markov processes on the circle, which is a natural extension of the results given in \citet{holzmann2006hidden, abe2017circular}.

Although \cite{harvey2024modelling} pointed out the sampling property of CACF under i.i.d. assumptions, their definition of the CACF is not the same as ours. The asymptotic properties of the sample CACF/CPACF are important; however, these remain future research topics.

In what follows, we illustrate the CACF for the MTD-AR($p$) models for $p=1,2$.
\begin{rem}[MTD-AR(1) case] From (\ref{eq:recursive_relation}), we have
\begin{align*}
\Gamma_k = D_{1}Q_1 \Gamma_{k-1} = (D_{1}Q_1)^k \Gamma_{0}~~~
\textrm{and}~~~
\det \Gamma_k = (q_1\rho_1^{2})^k/4.
\end{align*}
Then, we observe $r^{(C)}_k = q_1^k \rho_1^{2k}$. This is the same as the result of \citet{abe2017circular}. If we choose $g$ as the von Mises distribution with location $\mu_{\textrm{VM}}$ and concentration $\kappa_{\textrm{VM}}$, then $r^{(C)}_k = q^k (\frac{I_1(\kappa_{\textrm{VM}})}{I_0(\kappa_{\textrm{VM}})})^{2k}$, where $I_{\nu}(x)$ is the modified Bessel function of the first kind of order $\nu$. 
\end{rem}
\begin{rem}[MTD-AR(2) case] To solve the general expression for $\Gamma_k$ using the recursive expression 
\begin{align*}
\Gamma_k = a_1D_{1}Q_1\Gamma_{k-1} + a_2D_{1}Q_2 \Gamma_{k-2},
\end{align*}
we observe for $k=1$ and $\Gamma_0= I_2/2$ that
\begin{align*}
\Gamma_1 = \frac{a_1}{2} D_{1}Q_1 I_2 + a_2D_{1}Q_2 \Gamma_{-1}.
\end{align*}
Recall that $ \Gamma_{-1} =  \Gamma_{1}^{T}$ and writing $\Gamma_{1}= [\gamma_{1,ij}]_{i,j=1,2}$, the entries $\gamma_{1, ij}$ are the solution of the following equations:
\begin{align*}
&\begin{pmatrix}
1-a_2\rho_1\cos \mu_1 & a_2\rho_1q_2\sin \mu_1 & 0 & 0 \\
0 & 1 & -a_2\rho_1 \cos\mu_1 &  a_2\rho_1 q_2\sin \mu_1  \\
-a_2\rho_1\sin\mu_1 & -a_2\rho_1q_2\cos\mu_1 & 1 & 0 \\
0 & 0 & -a_2\rho_1\sin\mu_1 & 1-a_2\rho_1q_2 \cos\mu_1
\end{pmatrix}
\begin{pmatrix}
\gamma_{1,11} \\ \gamma_{1,12} \\ \gamma_{1,21} \\ \gamma_{1,22}
\end{pmatrix}
=
\frac{a_1 \rho_1}{2} 
\begin{pmatrix}
\cos \mu_1 \\ -q_1 \sin \mu_1\\  \sin \mu_1 \\ q_1 \cos \mu_1
\end{pmatrix}.
\end{align*}
Then, the CACF can be calculated from (\ref{eq:recursive_relation}).
\end{rem}

As the expression for the CACF for MTD-AR($p$) becomes complicated, we assume the following on the binding density, which makes the model simple and concise.

\begin{ass}\label{ass:mu=0}
The mean direction for the binding density $g$ is fixed at $\mu_1 =0$.
\end{ass}

This assumption implies that $\cos \Theta_t$ and $\sin \Theta_t$ are independent of each other, as can be convinced from the definition of $D_1$ whose off-diagonal elements became 0. In addition, this assumption ensures the symmetry of the covariance matrices $\Gamma_j$ such that $\Gamma_{-j} = \Gamma_j^{T} = \Gamma_j$.

\begin{rem}[MTD-AR(2) case (ii)] Assume that Assumptions 1 and 2 hold. Then, the expression (\ref{eq:recursive_relation})
reduces to 
\begin{align*}
\Gamma_1 =
 \begin{pmatrix}
1 -a_2 \rho_1 & 0 \\ 0 & 1- a_2 \rho_1 q_2
\end{pmatrix}^{-1}
 \cdot \frac{1}{2}
 \begin{pmatrix}
a_1\rho_1 & 0 \\ 0 &  a_1 \rho_1 q_1
\end{pmatrix}
\end{align*}
and
\begin{align*}
\det \Gamma_1  = \frac{ q_1 (a_1\rho_1)^2 }{4(1 -a_2 \rho_1 )(1- a_2 \rho_1 q_2) }.
\end{align*}
Similarly, for $k=2$,
\begin{align*}
\Gamma_2 = 
\frac{1}{2}
\begin{pmatrix}
\frac{(a_1\rho_1)^2}{1-a_2\rho_1} +a_2\rho_1 & 0 \\
0 & \frac{(q_1 a_1\rho_1)^2}{1-q_2a_2\rho_1} +q_2a_2\rho_1 
\end{pmatrix}
\end{align*}
and 
\begin{align*}
\det \Gamma_2 = 
\frac{1}{4}
\left(
\frac{(a_1\rho_1)^2}{1-a_2\rho_1} +a_2\rho_1  
\right)
\left(
 \frac{(q_1 a_1\rho_1)^2}{1-q_2a_2\rho_1} +q_2a_2\rho_1 
\right).
\end{align*}
Write $\Gamma_k = [ \gamma_{k, ij}]_{i,j=1,2}$, then we have
\begin{align*}
\det \Gamma_k =&  \gamma_{k,11} \gamma_{k,22}
= 
(a_1\rho_1 \gamma_{k-1,11} + a_2\rho_1 \gamma_{k-2,11})
(q_1a_1\rho_1 \gamma_{k-1,22} + q_2 a_2\rho_1 \gamma_{k-2,22}).
\end{align*}
Then, the lag $k$ CACF $r^{(C)}_k$s are calculated as follows:
\begin{align*}
 r^{(C)}_1 =& \frac{q_1 (a_1\rho_1)^2 }{(1-a_2\rho_1)(1-q_2 a_2\rho_1) },\\
 r^{(C)}_2 = &
\left(
\frac{(a_1\rho_1)^2}{1-a_2\rho_1} +a_2\rho_1  
\right)
\left(
 \frac{(q_1 a_1\rho_1)^2}{1-q_2a_2\rho_1} +q_2a_2\rho_1 
\right),\\
r^{(C)}_k =& 
4(a_1\rho_1 \gamma_{k-1,11} + a_2\rho_1 \gamma_{k-2,11})
(q_1a_1\rho_1 \gamma_{k-1,22} + q_2 a_2\rho_1 \gamma_{k-2,22})
~~~\textrm{for}~~k\geq 3.
\end{align*}
\end{rem}
To evaluate CACF for a general MTD-AR($p$) process, let
\begin{align}
\phi_1(z) =  1- a_1\rho_1 z - \cdots  - a_p  \rho_1 z^{p} 
\label{eq:polynomial1}
\end{align}
and
\begin{align}
\phi_2(z) = 1- q_1 a_1\rho_1 z - \cdots  - q_p a_p  \rho_1 z^{p} 
\label{eq:polynomial2}
\end{align}
 be the characteristic polynomials for $\gamma_{k,11}$ and $\gamma_{k,22}$, respectively. We see the CACF $r_{k}^{(C)}=4 \gamma_{k, 11}\gamma_{k, 22} $ is determined by the product of the difference equations for $k>0$
\begin{align}
\phi_1(B)\gamma_{k, 11} =0 ~~~\textrm{and}~~~ \phi_2(B)\gamma_{k, 22}   = 0 
\label{eq:arp}
\end{align}
where $B$ denotes the backshift operator. 
Then, we obtain
\begin{align*}
\phi_1(z)  \phi_2(z)= 
\prod_{i=1}^{m_1} (1 - G_{1,i} z)^{d_{1,i}} 
\prod_{i=1}^{m_2} (1 - G_{2,i} z)^{d_{2,i}},
\end{align*}
where $G_{1,i}^{-1}~(i=1,2,\ldots, m_1)$ and $G_{2,i}^{-1}~(i=1,2,\ldots, m_2)$ are the roots of the polynomials $\phi_{1}(z) =0$ and $\phi_{2}(z) =0$, respectively, and $\sum_{i=1}^{m_j} d_{j,i}=p$ for $j=1,2$. Then, the CACF is given by the following result.

\begin{thm}\label{th:2}
 We consider that Assumptions 1 and 2 hold. The lag $k$ CACF for MTD-AR($p$) process is given by
\begin{align*}
r^{(C)}_k =
\left(
 \sum_{i=1}^{m_1} G_{1,i}^k \sum_{j=0}^{d_{1,i}-1} A_{1,ij} k^j
 \right)
 \left(
 \sum_{i=1}^{m_2} G_{2,i}^k \sum_{j=0}^{d_{2,i}-1} A_{2, ij} k^j
 \right)
 .
 \end{align*}
 If $d_{j,i}=1$ for all $i=1,\ldots, p$ and $j=1,2$, $G_{1,i}^{-1}$ and $G_{2,i}^{-1}$ are all distinct and the CACF reduces to 
 \begin{align*}
 r^{(C)}_k = \left( \sum_{i=1}^p A_{1,i} G_{1,i}^k \right)  \left( \sum_{i=1}^p A_{2,i} G_{2,i}^k \right), ~~k>0,
 \end{align*}
 where $A_{1,i}$ and $A_{2,i}$ are the constants obtained recursively using (\ref{eq:recursive_relation}) for $k=1,2,\ldots, p$.
 \end{thm}
 
\subsection{Circular partial autocorrelation functions}\label{sec:3.2}

So far, we have investigated second-order stationarity and CACF of the MTD-AR($p$) models. Here, we provide a CPACF for MTD-AR($p$) models. The process considered here is the bivariate process $\bm{U}_t=(\cos \Theta_{t}, \sin \Theta_{t})^T$ defined in Section~{\ref{sec:3.1}}. Recall that as the expressions of the forward and backward innovations of $\bm{U}_t$ are different, the simple expression of the partial autocorrelation function (PACF) is difficult to obtain.  See \cite{morf1978covariance} and \cite{degerine1990canonical} for PACF for multivariate time series.

 We now define the CPACF following \cite{degerine1990canonical}. The best linear predictor $\widehat{\bm{U}}_{t,s}$ is the projection onto the $s$ past valued linear space. We write the $s$-th order forward innovation $\bm{\varepsilon}_t$ as
\begin{align*}
\bm{\varepsilon}_{t,s} = \bm{U}_t-\widehat{\bm{U}}_{t,s}
=\bm{U}_t - \sum_{j=1}^s \Xi_{j}\bm{U}_{t-j}, 
\end{align*} 
The solution of $\Xi_{j}$, $j=1,\ldots, s$ is obtained by the following linear equations:
\begin{align}
\begin{bmatrix}
\Xi_{1} &  \Xi_{2} &\cdots & \Xi_{s}
\end{bmatrix}
\begin{bmatrix}
\Gamma_0 & \Gamma_{1} & \cdots & \Gamma_{s-1} \\
\Gamma_{-1} & \Gamma_{0} & \cdots & \Gamma_{s-2} \\
\vdots & \vdots      & \ddots       & \vdots \\
\Gamma_{1-s} & \Gamma_{2-s} & \cdots & \Gamma_0    
\end{bmatrix}
=
\begin{bmatrix}
\Gamma_1 & \Gamma_2 & \cdots & \Gamma_s
\end{bmatrix}.
\label{forward}
\end{align}
Similarly, the $n$-th order backward innovation $\bm{\varepsilon}_{t,s}^*$ can be defined as 
\begin{align*}
\bm{\varepsilon}_{t,s}^* = \bm{U}_t-\widehat{\bm{U}}^*_{t,s}
=  \bm{U}_t -\sum_{j=1}^s \Xi_{j}^*\bm{U}_{t+j}.
\end{align*}  
The solution of $\Xi_{j}^*$, $j=1,\ldots, s$ is obtained by following linear equations:
\begin{align}
\begin{bmatrix}
  \Xi_{1}^* &   \Xi_{2}^* &\cdots & \Xi_{s}^*
\end{bmatrix}
\begin{bmatrix}
\Gamma_0 & \Gamma_{-1} & \cdots & \Gamma_{1-s} \\
\Gamma_1 & \Gamma_{0} & \cdots & \Gamma_{2-s} \\
\vdots & \vdots & \ddots
& \vdots \\
\Gamma_{s-1} & \Gamma_{s-2} & \cdots & \Gamma_0    
\end{bmatrix}
=
\begin{bmatrix}
\Gamma_{-1} &\Gamma_{-2}& \cdots & \Gamma_{-s}
\end{bmatrix}.
\label{backward}
\end{align}
Then, the partial autocovariance matrix of the process $U_t$ is defined by
\begin{align*}
\Psi_s = E [ \bm{\varepsilon}_{t,s-1}{{\bm{\varepsilon}_{t-s,s-1}^{*}}^{T}}]. 
\end{align*} 
Similar to the definition of the CACF, we propose a CPACF as 
\begin{align}
\psi_s^{(C)}=  \frac{\textrm{det}(\Psi_s)}{ \{ \textrm{det}E(\bm{\varepsilon}_{t,s-1}
\bm{\varepsilon}_{t,s-1}^{T})
\textrm{det}E(\bm{\varepsilon}_{t,s-1}^*{\bm{\varepsilon}_{t,s-1}^{*}}^{T}) \}^{1/2} }.
\label{CPACF}
\end{align}

The general expressions for the CPACF are complicated forms; however, we obtain rather simple expressions under Assumption 2. This is because the forward and backward innovations are equivalent due to the symmetry of matrices $\Gamma_k$.
The CPACF is stated in the following theorem.

\begin{thm}\label{th:4}
Assume that Assumptions 1 and 2 hold. Then, the lag $s$ CPACF of the MTD-AR(p) process $ \{\Theta_t \}_{t \in \mathbb{Z}} $ given by (\ref{CPACF})  can be obtained by solving
\begin{align}
\label{eq:CPACF(det form)}
\psi_{s}^{(C)}
=
\left.
\begin{vmatrix}
 \bm{\Gamma}_0 & \bm{\Gamma}_1& \cdots & \bm{\Gamma}_{s-2} & \bm{\Gamma}_{1} \\
  \bm{\Gamma}_1 & \bm{\Gamma}_0 & \cdots &  \bm{\Gamma}_{s-3} &  \bm{\Gamma}_{2} \\
  \vdots & \vdots & & \vdots & \vdots \\ 
  \bm{\Gamma}_{s-1} & \bm{\Gamma}_{s-2} & \cdots &  \bm{\Gamma}_{1} &  \bm{\Gamma}_{s} 
 \end{vmatrix}
 \middle/
\begin{vmatrix}
 \bm{\Gamma}_0 & \bm{\Gamma}_1& \cdots & \bm{\Gamma}_{s-2} & \bm{\Gamma}_{s-1} \\
  \bm{\Gamma}_1 & \bm{\Gamma}_0 & \cdots &  \bm{\Gamma}_{s-3} &  \bm{\Gamma}_{s-2} \\
  \vdots & \vdots & & \vdots & \vdots \\ 
  \bm{\Gamma}_{s-1} & \bm{\Gamma}_{s-2} & \cdots &  \bm{\Gamma}_{1} &  \bm{\Gamma}_0
 \end{vmatrix}.
 \right.
\end{align}
\end{thm}

The sample CPACF is defined by
\begin{align*}
\widehat{\psi}_k^{(C)}
=
\left.
\begin{vmatrix}
 \widehat{\bm{\Gamma}}_0 &  \widehat{\bm{\Gamma}}_1& \cdots &  \widehat{\bm{\Gamma}}_{k-2} &  \widehat{\bm{\Gamma}}_{1} \\
   \widehat{\bm{\Gamma}}_1^T &  \widehat{\bm{\Gamma}}_0 & \cdots &   \widehat{\bm{\Gamma}}_{k-3} &   \widehat{\bm{\Gamma}}_{2} \\
  \vdots & \vdots & & \vdots & \vdots \\ 
   \widehat{\bm{\Gamma}}_{k-1}^T &  \widehat{\bm{\Gamma}}_{k-2}^T & \cdots &   \widehat{\bm{\Gamma}}_{1}^T &   \widehat{\bm{\Gamma}}_{k} 
 \end{vmatrix}
 \middle/
 \begin{vmatrix}
  \widehat{\bm{\Gamma}}_0 &  \widehat{\bm{\Gamma}}_1& \cdots &  \widehat{\bm{\Gamma}}_{k-2} &  \widehat{\bm{\Gamma}}_{k-1} \\
  \widehat{\bm{\Gamma}}_1^T &  \widehat{\bm{\Gamma}}_0 & \cdots &   \widehat{\bm{\Gamma}}_{k-3} &   \widehat{\bm{\Gamma}}_{k-2} \\
  \vdots & \vdots & & \vdots & \vdots \\ 
   \widehat{\bm{\Gamma}}_{k-1}^T &  \widehat{\bm{\Gamma}}_{k-2}^T & \cdots &   \widehat{\bm{\Gamma}}_{1}^T &   \widehat{\bm{\Gamma}}_0
 \end{vmatrix}. \right.
\end{align*}
where $\widehat{\bm{\Gamma}}_k$ is defined by (\ref{eq:samplecovarinace2}). It must be noted that the sample covariance matrix at lag $k$ of $\widehat{\bm{\Gamma}}_k$ is not a symmetric matrix.

\begin{rem}[MTD-AR(2) case]
 Here, we provide the CPACF of the MTD-AR(2) process as follows. For $k=1$, we observe 
 \begin{align*}
\Gamma_1 =
 \begin{pmatrix}
1 -a_2 \rho_1 & 0 \\ 0 & 1- a_2 \rho_1 q_2
\end{pmatrix}^{-1}
 \cdot \frac{1}{2}
 \begin{pmatrix}
a_1\rho_1 & 0 \\ 0 &  a_1 \rho_1 q_1
\end{pmatrix}
\end{align*}
and
\begin{align*}
\det \Gamma_1  = \frac{ q_1 (a_1\rho_1)^2 }{4(1 -a_2 \rho_1 )(1- a_2 \rho_1 q_2) }.
\end{align*}
Then, $\psi_1^{(C)}$ becomes
\begin{align*}
\psi_1^{(C)} = &
\frac{ q_1 (a_1\rho_1)^2 }{(1 -a_2 \rho_1 )(1- a_2 \rho_1 q_2) } 
=
\frac{a_1\rho_1}{1 -a_2 \rho_1}\cdot \frac{q_1a_1\rho_1}{1- a_2 \rho_1 q_2}\\
=:& \rho_c(1) \cdot \rho_s(1),
\end{align*}
where $\rho_c(1)$ and $\rho_s(1)$ are the lag 1 partial autocorrelation function of the $\cos \Theta_t$ and $\sin \Theta_t$ processes whose characteristic equations are defined by equations (\ref{eq:polynomial1}) and (\ref{eq:polynomial2}), respectively, with order $p=2$. Similarly, for $k=2$, we have
\begin{align*}
\psi_2^{(C)} 
= 
\left. 
\begin{vmatrix}
\bm{\Gamma}_0 & \bm{\Gamma}_1 \\
\bm{\Gamma}_1 & \bm{\Gamma}_2 
\end{vmatrix}
\middle/
\begin{vmatrix}
\bm{\Gamma}_0 & \bm{\Gamma}_1 \\
\bm{\Gamma}_1 & \bm{\Gamma}_0 
\end{vmatrix}.
\right.
\end{align*}
After some calculations, the denominator becomes
\begin{align*}
\begin{vmatrix}
\bm{\Gamma}_0 & \bm{\Gamma}_1 \\
\bm{\Gamma}_1 & \bm{\Gamma}_0 
\end{vmatrix}=
 \frac{1}{4}(1- \rho_c(1)^2) \times\frac{1}{4}(1- \rho_s(1)^2).
\end{align*}
As for the numerator, we obtain
\begin{align*}
\begin{vmatrix}
\bm{\Gamma}_0 & \bm{\Gamma}_1 \\
\bm{\Gamma}_1 & \bm{\Gamma}_2 
\end{vmatrix}=
 \frac{1}{4}(\rho_c(2)- \rho_c(1)^2) \times\frac{1}{4}(\rho_s(2)- \rho_s(1)^2),
\end{align*}
where $\rho_c(2)= \frac{(a_1\rho_1)^2}{1-a_1\rho_1} +a_2\rho_1 $ and $\rho_s(2)=\frac{(q_1 a_1\rho_1)^2}{1-q_2a_1\rho_1 } +q_2a_2\rho_1$ are the lag 2 partial autocorrelation function for the AR(2) of the $\cos \Theta_t$ and $\sin \Theta_t$ processes. Then, we get
\begin{align*}
\psi_2^{(C)}=
\frac{ \rho_c(2)- \rho_c(1)^2}{1- \rho_c(1)^2}
\times
\frac{ \rho_s(2)- \rho_s(1)^2}{1- \rho_s(1)^2}.
\end{align*}
This indicates that the CPACF at lag 2 is the product of the partial autocorrelation function at lag 2 for the linear AR(2) process of $\cos \Theta_t$ and $\sin \Theta_t$. Finally, for $k=3$, we have
 \begin{align*}
\psi_3^{(C)}
=
\left.
\begin{vmatrix}
\bm{\Gamma}_0 & \bm{\Gamma}_1 & \bm{\Gamma}_1 \\
\bm{\Gamma}_1 & \bm{\Gamma}_0 & \bm{\Gamma}_2\\
\bm{\Gamma}_2 & \bm{\Gamma}_1 & \bm{\Gamma}_3
\end{vmatrix}
\middle/
\begin{vmatrix}
\bm{\Gamma}_0 & \bm{\Gamma}_1 & \bm{\Gamma}_2  \\
\bm{\Gamma}_1 & \bm{\Gamma}_0 & \bm{\Gamma}_1 \\
\bm{\Gamma}_2 & \bm{\Gamma}_1 & \bm{\Gamma}_0
\end{vmatrix}
.\right.
\end{align*}
The numerator becomes
\begin{align*} 
& (1/2)^3 \{
(\rho_c(3) -\rho_c(1)\rho_c(2)) -\rho_c(1)^2(\rho_c(3)-\rho_c(1) ) 
+\rho_c(2)\rho_c(1)(\rho_c(2) -1)  
\}\\
\times &
(1/2)^3 \{
(\rho_s(3) -\rho_s(1)\rho_s(2)) 
-\rho_s(1)^2(\rho_s(3)-\rho_s(1) ) +\rho_s(2)\rho_s(1)(\rho_s(2) -1)  
\},
\end{align*} 
where $\rho_c(3)=a_1\rho_1 \rho_c(2) + a_2\rho_1 \rho_c(1)$ and $\rho_s(3)=
q_1a_1\rho_1 \rho_s(2) + q_2 a_2\rho_1 \rho_s(1)$ are the partial autocorrelation function for the AR(2) of $\cos\Theta_t$ and $\sin\Theta_t$ 
processes, respectively. As we have $\rho_c(3)=\rho_s(3)=0$, the CPACF at lag 3 becomes 0. The analogous calculation yields that $\psi_k^{(C)} =0$ for $k\geq 4$.
\end{rem}

\subsection{Spectral density function of the higher-order circular time series}\label{sec:3.3}

The quantity $\det \Gamma_k$ can be considered as the circular autocovariance function as explained in \citet{abe2017circular}. Hence, by using the Fourier series expansion of $\det \Gamma_k$, we can obtain the spectral density for the circular stationary process. 
We call $ \det \Gamma_k $ in (\ref{eq:CACF})  the valid circular autocovariance function, and the corresponding spectral density is expressed as
\begin{align*}
f_{\Theta}(\omega) = \frac{1}{2\pi} \sum_{k=-\infty}^{\infty} e^{-\mathrm{i} \omega k} \det \Gamma_k
=
\frac{1}{2\pi} \sum_{k=-\infty}^{\infty} e^{-\mathrm{i} \omega k}  \begin{vmatrix}
\gamma_{k,11} & \gamma_{k,12} \\
\gamma_{k,21} & \gamma_{k,22}
\end{vmatrix}\red{,}
\end{align*}   
and the inverse Fourier transforms for all $k \in \mathbb{Z}$,
\begin{align*}
\det \Gamma_k
= \frac{1}{2\pi} \int_{-\pi}^{\pi} e^{\mathrm{i}k \omega}
f_{\Theta}(\omega) d\omega.
\end{align*}
Then, we obtain the following result.

\begin{thm}\label{th:3}
 Assume that Assumptions 1 and 2 hold. Let the AR characteristic polynomials $\phi_1(z)$ and $\phi_2(z)$ of the processes $\cos \Theta_t$ and $\sin \Theta_t$ be defined by (\ref{eq:polynomial1}) and (\ref{eq:polynomial2}), respectively. Then, the spectral density function for MTD-AR($p$) process defined by (\ref{eq:MTD}) is given by
\begin{align*}
f_{\Theta}(\omega) =\int_{-\pi}^{\pi}  
 f_{X_1}(\lambda) f_{X_2}(\omega- \lambda) d\lambda,
\end{align*}  
where $f_{X_1}(\omega)$ and $f_{X_2}(\omega)$ are the spectral density functions of the process $\cos\Theta_t$ and $\sin\Theta_t$, which are  given by
\begin{align*}
f_{X_1}(\omega) =
 \frac{ 1-2 \rho_1\sum_{i=1}^{p} a_i \gamma_{i,11}}{4\pi  | \phi_1(e^{-\mathrm{i}\omega})|^2}~~~\text{and}~~~
f_{X_2}(\omega) =
 \frac{ 1-2 \rho_1\sum_{i=1}^{p} q_ia_i \gamma_{i,22}}{4\pi  | \phi_2(e^{-\mathrm{i}\omega})|^2},
\end{align*}
respectively.
\end{thm}

Practically, the spectral density is calculated as follows: 
\begin{align*}
&f_{\Theta}(\omega) 
=
\int_{-\pi}^{\pi}  
 f_{X_1}(\lambda)  
 f_{X_2}(\omega- \lambda) d\lambda \\
=&\frac{ (1-2 \rho_1\sum_{i=1}^{p} a_i \gamma_{i,11})(1-2 \rho_1\sum_{i=1}^{p} q_ia_i \gamma_{i,22})}{16\pi^2} \\
\times& \int_{-\pi}^{\pi}
\frac{d\lambda}{\prod_{i=1}^{m_1} (1 - G_{1,i} e^{-\mathrm{i}\lambda})^{d_{1,i}} 
(1 - G^\ast_{1,i} e^{\mathrm{i}\lambda})^{d_{1,i}}
\prod_{i=1}^{m_2} (1 - G_{2,i} e^{-\mathrm{i}(\omega-\lambda)})^{d_{2,i}}
(1 - G^\ast_{2,i} e^{\mathrm{i}(\omega-\lambda)})^{d_{2,i}}
}.
 \end{align*}
Setting $z=e^{\mathrm{i}\lambda}$, we have $d\lambda=dz/(\mathrm{i}z)$, and the above integral becomes
\begin{align}\label{eq:integral part}
&\frac{1}{\textrm{i}}
\int_{C}
\frac{dz}{z\prod_{i=1}^{m_1} (1 - \frac{G_{1,i}}{z} )^{d_{1,i}} 
(1 - G^\ast_{1,i} z)^{d_{1,i}}
\prod_{i=1}^{m_2} (1 - G_{2,i} e^{-\mathrm{i}\omega}z)^{d_{2,i}}
(1 - \frac{G^\ast_{2,i} e^{\mathrm{i}\omega}}{z})^{d_{2,i}}
} \nonumber \\
=&\frac{1}{\textrm{i}}
\int_{C}
\frac{1}{z}
\frac{1}{\prod_{i=1}^{m_1} (\frac{1}{z})^{d_{1,i}}(z - G_{1,i} )^{d_{1,i}} 
(-G^\ast_{1,i})^{d_{1,i}}(z-\frac{1}{G^\ast_{1,i}}  )^{d_{1,i}}} \nonumber \\
& \qquad \times
\frac{1}{\prod_{i=1}^{m_2} (-G_{2,i} e^{-\mathrm{i}\omega})^{d_{2,i}}(z - \frac{1}{G_{2,i} e^{-\mathrm{i}\omega}})^{d_{2,i}}
(\frac{1}{z})^{d_{2,i}}(z - G^\ast_{2,i} e^{\mathrm{i}\omega})^{d_{2,i}}
} dz \nonumber \\
=&
\frac{1}{\textrm{i}
\prod_{i=1}^{m_1}(-G^\ast_{1,i})^{d_{1,i}}
\prod_{i=1}^{m_2} (-G_{2,i} e^{-\mathrm{i}\omega})^{d_{2,i}}} \nonumber \\
&\times \int_{C}
\frac{z^{2p-1}}{\prod_{i=1}^{m_1} (z - G_{1,i} )^{d_{1,i}} 
(z-\frac{1}{G^\ast_{1,i}}  )^{d_{1,i}}\prod_{i=1}^{m_2} (z - \frac{1}{G_{2,i} e^{-\mathrm{i}\omega}})^{d_{2,i}}
(z - G^\ast_{2,i} e^{\mathrm{i}\omega})^{d_{2,i}}
} dz,
\end{align}
where $C=\{z \in \mathbb{C} : |z|=1 \}$.
The integrand in (\ref{eq:integral part}) has poles of order $d_{1,i}$ at $z=G_{1,i}$  $(i=1,2,\ldots,m_1)$, order $d_{2,i}$ at $z=G^\ast_{2,i} e^{\mathrm{i}\omega}$  $(i=1,2,\ldots,m_2)$. From the residue theorem, the integral is equal to $(2 \pi i)$ times the sum of all residues. 
Therefore, the spectral density of $\Theta_t$ is 
\begin{align}
\label{eq:SDF}
f_{\Theta}(\omega)=
&\frac{ (1- 2\rho_1\sum_{i=1}^{p} a_i \gamma_{i,11})(1- 2\rho_1\sum_{i=1}^{p} q_ia_i \gamma_{i,22})}{8\pi
\prod_{i=1}^{m_1}(-G^\ast_{1,i})^{d_{1,i}}
\prod_{i=1}^{m_2} (-G_{2,i} e^{-\mathrm{i}\omega})^{d_{2,i}}} 
\left\{
\sum_{i=1}^{m_1} \mathrm{Res}(G_{1,i})+ \sum_{i=1}^{m_2} \mathrm{Res}(G^\ast_{2,i} e^{\mathrm{i}\omega})
\right\}
\end{align}
\begin{rem}[MTD-AR(1) case]
Consider that the transition density in (\ref{eq:MTD}) is
$g(\theta_t-q_1 \theta_{t-1})$. We set the mean resultant length and mean resultant direction of $g$ are $\rho_1$ and $0$, respectively.
Then, we have $2\gamma_{1,11}=G_{1,1}=\rho_1$ and  $2\gamma_{1,22}=G_{2,1}=q_1 \rho_1$. 
By (\ref{eq:SDF}), the spectral density function is
\begin{align*}
f_\Theta(\omega)=\frac{ (1- \rho_1^2)^2}{8\pi
 q_1\rho_1^2 e^{-\mathrm{i}\omega}} 
\left\{
 \mathrm{Res}(\rho_1)+  \mathrm{Res}(q_1\rho_1e^{\mathrm{i}\omega})
\right\}.
\end{align*}
Here, 
\begin{align*}
\mathrm{Res}(\rho_1)=
\frac{\rho_1}{ 
(\rho_1-\frac{1}{\rho_1}  )
(\rho_1 - \frac{1}{q_1\rho_1 e^{-\mathrm{i}\omega}})
(\rho_1 - q_1\rho_1 e^{\mathrm{i}\omega})
}
=
\frac{q_1\rho_1^2 e^{-\mathrm{i}\omega}}
{(1-\rho_1^2)(1-q_1\rho_1^2 e^{-\mathrm{i}\omega})
(1-q_1e^{\mathrm{i}\omega})}
\end{align*}
and
\begin{align*}
\mathrm{Res}(q_1\rho_1e^{\mathrm{i}\omega})
=&
\frac{q_1\rho_1e^{\mathrm{i}\omega}}
{(q_1\rho_1e^{\mathrm{i}\omega} - \rho_1) 
(q_1\rho_1e^{\mathrm{i}\omega}-\frac{1}{\rho_1}  )
(q_1\rho_1e^{\mathrm{i}\omega} - \frac{1}{q_1\rho_1 e^{-\mathrm{i}\omega}})
}\\
=&\frac{-\rho_1^2}{(1-\rho_1^2)(1-q_1\rho_1^2 e^{\mathrm{i}\omega})
(1-q_1e^{\mathrm{i}\omega})}.
\end{align*}
Finally, we have 
\begin{align*}
f_\Theta(\omega)=
\frac{ (1- \rho_1^2)(1+\rho_1^2)}
{8\pi
 |1-q_1\rho_1^2e^{\mathrm{i}\omega}|^2
} 
.
\end{align*}
\end{rem}

\section{Maximum Likelihood Estimation}\label{sec:4}

For the MTD models studied by, for example, \cite{le1996modeling, raftery1994estimation}, the maximum likelihood estimation with EM algorithm is a standard estimation method, as it works well for a very large and highly nonlinear class of MTD models \cite{berchtold2002mixture}. However, the estimates of EM algorithms tend to have a local optimum solution rather than a global one. In addition, the iterative procedures need a large computational time to converge. This section provides the maximum likelihood estimation methods for unknown model parameters without implementing EM algorithms. Let the parameter vector be $\bm{\eta}=(a_1, \ldots, a_{p-1}, \rho)^T$ and $\bm{q} =(q_1, \ldots, q_p)^T$ for the MTD model in (\ref{eq:MTD}). The parameter space for $(a_1,\ldots, a_{p-1})$ is denoted by $\bm{A}$, which is defined as 
\begin{align*}
\bm{A} = \left\{ (a_1,\ldots, a_{p-1}) | 0 \leq a_i \leq 1-\delta_{A_1}, \sum_{i=1}^{p-1}a_i \leq 1-\delta_{A_2} \right\},
\end{align*}
for some $\delta_{A_1} >0$ and $\delta_{A_2}>0$. The parameter space of a concentration parameter of a binding density $g$ is denoted by $R_{\rho}=(0,\bar{R}_{\rho}]$. For the wrapped Cauchy distributions, we set $\rho \in [\delta_{\rho},1-\delta_{\rho}]$, and for the von Mises distribution, we set $\rho=\kappa$ and $\kappa \in [\delta_{\kappa},\bar{R}_{\kappa}]$, where $\delta_{\rho}$ and $\delta_{\kappa}$ are small constants and $0< \bar{R}_{\kappa}<\infty$. Then, we define a parameter space of $\bm{\eta}$ as $\bm{H}=\bm{A} \times R_{\rho}$. As the direction of the dependency parameter vector $\bm{q}$ is discrete values, the parameter spaces for $\bm{q}$ is denoted by $\bm{Q}$, which is defined as 
\begin{align*}
\bm{Q}= \left\{
(q_1, q_2,\ldots, q_p) | q_i\in\{1, -1\}, ~i=1,\ldots, p
\right\}.
\end{align*}
The log-likelihood function of the model becomes:
\begin{align*}
\ell_n(\bm{\eta}, \bm{q}) = \sum_{t=p + 1}^{n}  \ln
\left(
\sum_{i=1}^p a_i g(\theta_t -q_i \theta_{t-i}; \rho)
\right).
\end{align*}
 The MLE of $\hat{\bm{\eta}}_n$ and $\hat{\bm{q}}_n$ are obtained by maximizing the log-likelihood function over the parameter spaces $\bm{H}$ and $\bm{Q}$ as follows:
\begin{align*}
(\hat{\bm{\eta}}_n, \hat{\bm{q}}_n )=  \argmax_{ \bm{\eta}\in \bm{H}, \bm{q}\in \bm{Q}} \ell_n(\bm{\eta} , \bm{q}).
\end{align*}
We do not assume that the true binding density $g$ and true AR order $p$ belong to the set of distributions of a fitted model. Hence, we call the estimator defined above a quasi-likelihood estimator under the misspecified model. Hereafter, we assume that the direction of the dependency parameter vector $\bm{q}$ is prespecified for simplicity; we denote it by $\bm{q}_0=(q_{10},\ldots, q_{p0})$. In addition, the autoregressive order $p$ is also prespecified. To derive asymptotic properties of $\hat{\bm{\eta}}_n$ with known $\bm{q}_0$, we denote the MLE of $\bm{\eta}$ as follows:
\begin{align*}
\hat{\bm{\eta}}_n
= \argmax_{\bm{\eta}\in \bm{H}} \ell_n(\bm{\eta}, \bm{q}_0).
\end{align*} 
We denote $\bm{\eta}_0$ by
\begin{align*}
\bm{\eta}_0 = \mathop{\textrm{argmax}}_{\bm{\eta}\in \bm{H}}E\left[
\ln \left(
\sum_{i=1}^p a_i g(\Theta_t -q_{i0} \Theta_{t-i}; \rho)
\right)
 \right].
\end{align*}
We assume the following condition:

\begin{ass}
\label{ass3}
\begin{itemize}
\item[(a)]  $\bm{H}$ is compact, and $\bm{\eta}_0$ belongs to $\bm{H}$.
\item[(b)] The binding density $g(\cdot)$ is continuous. In addition, $g(\cdot)$ satisfies  
\begin{align*}
\sup_{\rho\in R_{\rho}}g(\theta; \rho) < \overline{M} ~~\textrm{and}~~
\inf_{\rho\in R_{\rho}}g(\theta; \rho) > \underline{M}
\end{align*} 
for some constants $\overline{M}>0$ and $\underline{M}>0$, for any $\theta\in [-\pi, \pi)$.
\item[(c)] The first two derivatives of the binding density $g(\cdot)$ with respect to parameters $\rho$ are continuous and bounded for all $\theta$ and $\rho\in R_{\rho}$.
\item[(d)] The Fisher information matrix \\$\bm{I}(\bm{\eta}_0) := - E_{\eta_0}\left[ (\partial^2/\partial \eta\partial \eta^T) \ln\left( 
\sum_{i=1}^p a_i g(\Theta_t -q_{i0} \Theta_{t-i}; \rho)\right) \right]$ is non-singular.
\item[(e)] The family $\{ \sum_{i=1}^p a_i g(\Theta_t -q_{i0} \Theta_{t-i}; \rho)\}$ is identifiable, which implies for parameters $\bm{\eta}_1=(a_{1,1},\ldots , 
a_{p-1,1},\rho_1 ) \ne
(a_{1,2},\ldots,a_{p-1,2},\rho_2 )=\bm{\eta}_2$, we observe $ \{ \sum_{i=1}^p a_{i,1} g(\Theta_t -q_{i0} \Theta_{t-i}; \rho_1)\} \ne \{ \sum_{i=1}^p a_{i,2} 
g(\Theta_t -q_{i0} \Theta_{t-i}; \rho_2)\}$.
\end{itemize}
\end{ass}

Assumptions \ref{ass3}(b) and (c) ensure that the Markov kernel function $\sum_{i=1}^p a_{i} g(\Theta_t -q_{i0} \Theta_{t-i}; \rho)$ is continuous and its logarithm is bounded for all $\bm{\eta}$. As we assume that $\inf_{\rho\in R_{\rho}}g(\theta; \rho) > \underline{M}$ in Assumption \ref{ass3}(b), this guarantees the integrability condition $E_{\eta_0}\{\ln (\sum_{i=1}^p a_{i} g(\Theta_t -q_{i0} \Theta_{t-i}; \rho))\} > -\infty$. In addition, Assumption \ref{ass3}(b) implies that the state space of a process $\{\Theta_t\}$ is finite or compact. It implies that the Markov kernel $p(\theta_t|\theta_{t-1}, \ldots, \theta_{t-p})$ is uniformly geometrically ergodic; see, for more details, \citep[Assumption A 13.1]{douc2014nonlinear}. Therefore, the initial distribution does not affect the stationary distribution of the MTD-AR($p$) process. Examples of binding density functions that satisfy Assumption \ref{ass3}(b) include well-known symmetric circular densities, such as the von Mises and wrapped Cauchy distributions with location $\mu=0$. On the contrary, the cardioid distribution and the Jones--Pewsey distribution with shape parameter $\psi=1$ should be eliminated from the candidates of the binding densities. 

We obtain the following consistency and asymptotic normality properties for the MLE.

\begin{thm}\label{th5} (Consistency)
Let $\{ \hat{\bm{\eta}}_n \}$ be a sequence of MLE and Assumptions 1--3 hold. Then, we have as $n\to\infty$,
\begin{align*}
\hat{\bm{\eta}}_n \mathop{\to}_{a.s.} \bm{\eta}_0.
\end{align*}
\end{thm}
 
 To obtain the limiting distribution of $\hat{\bm{\eta}}_n$, we assume the following instead of Assumption 3(a).
\\
\\
~
\noindent
\textbf{Assumption 3}
 \textbf{(a')}  $\bm{H}$ is convex and compact, and $\bm{\eta}_0$ belongs to the interior of $\bm{H}$.
\\
\\
~
 The following result gives the limiting distribution of $\hat{\bm{\eta}}_n$.
 
\begin{thm}\label{th6}(Asymptotic normality)
Under Assumptions 1, 2, 3(a'), and 3(b)--(e), we have 
\begin{align*}
\sqrt{n} (  \hat{\bm{\eta}}_n - \bm{\eta}_0) \mathop{\longrightarrow}_{d}
N(\bm{0}, I(\bm{\eta}_0)^{-1} \bm{J}(\bm{\eta}_0) I(\bm{\eta}_0)^{-1}),
\end{align*}
where $I(\bm{\eta}_0)$ is defined by Assumption 3(d) and $\bm{J}(\bm{\eta}_0)$ is defined by
\begin{align*}
J(\bm{\eta}_0) = E_{\bm{\eta}_0} \left[
\left(
\frac{\partial}{\partial \bm{\eta}}\ell_n(\bm{\eta},\bm{q}_0) \right)
\left(
\frac{\partial}{\partial \bm{\eta}}\ell_n(\bm{\eta},\bm{q}_0) \right)^T
 \right].
\end{align*}
\end{thm}
~
\\
\\
~
For a misspecified model, MLE strongly converges to the parameter vector that minimizes the Kullback-Leibler divergence between the true model and the postulated likelihood functions. If we consider a well-specified model, the generalized information matrix equality holds such that $\bm{I}(\bm{\eta}_0) = \bm{J}(\bm{\eta}_0)$, which leads the asymptotic distribution of the MLE as follows:
\begin{align*}
\sqrt{n} (\hat{\bm{\eta}}_n -\bm{\eta}_0)
\mathop{\to}_{d}
N\left( \bm{0},
\bm{I}(\bm{\eta}_0)^{-1} \right).
\end{align*}
The MLE of the direction of dependency parameters $\hat{\bm{q}}_n$ is obtained by following procedures. Let the $j$-th elements of the parameter space $\bm{Q}$ be denoted by $\bm{q}_j$~($j=1,\ldots, 2^{p}$). Then, $\hat{\bm{q}}_n$ is defined by
\begin{align*}
\bm{\hat{q}}_n = \mathop{\textrm{argmax}}_{q_j \in Q} \ell_n(\hat{\bm{\eta}}_n(\bm{q}_j), \bm{q}_j),
\end{align*} 
where $\hat{\bm{\eta}}_n(\bm{q}_j)$ is the MLE of $\bm{\eta}$ given $\bm{q}_j$, and the corresponding MLE based on $\hat{\bm{\eta}}_n$ is denoted by $\hat{\bm{\eta}}_n(\hat{\bm{q}}_n)$. 

The unknown order $p$ and the direction of dependency parameter $\bm{q}$ are selected using well-known information criteria, Akaike information criteria (AIC) and Bayesian information criteria (BIC).
The parameters of MTD-AR($p$) models consist of the mixing weights and the concentration of the binding density; hence, the number of parameters reduces to $p$. The definitions of the AIC and BIC are as follows:
\begin{align*}
\textrm{AIC}= - 2 \ell_n(\hat{\bm{\eta}}_n(\hat{\bm{q}}_n), \hat{\bm{q}}_n) +2p
\end{align*}
and
\begin{align*}
\textrm{BIC}= - 2 \ell_n(\hat{\bm{\eta}}_n(\hat{\bm{q}}_n), \hat{\bm{q}}_n) +p \log n.
\end{align*}

\section{Monte Carlo Simulations and Numerical Examples}\label{sec:5}

\begin{figure}[h]
 \begin{minipage}{0.45\hsize}
  \centering
  \includegraphics[width=.9\textwidth,height=0.15\textheight]{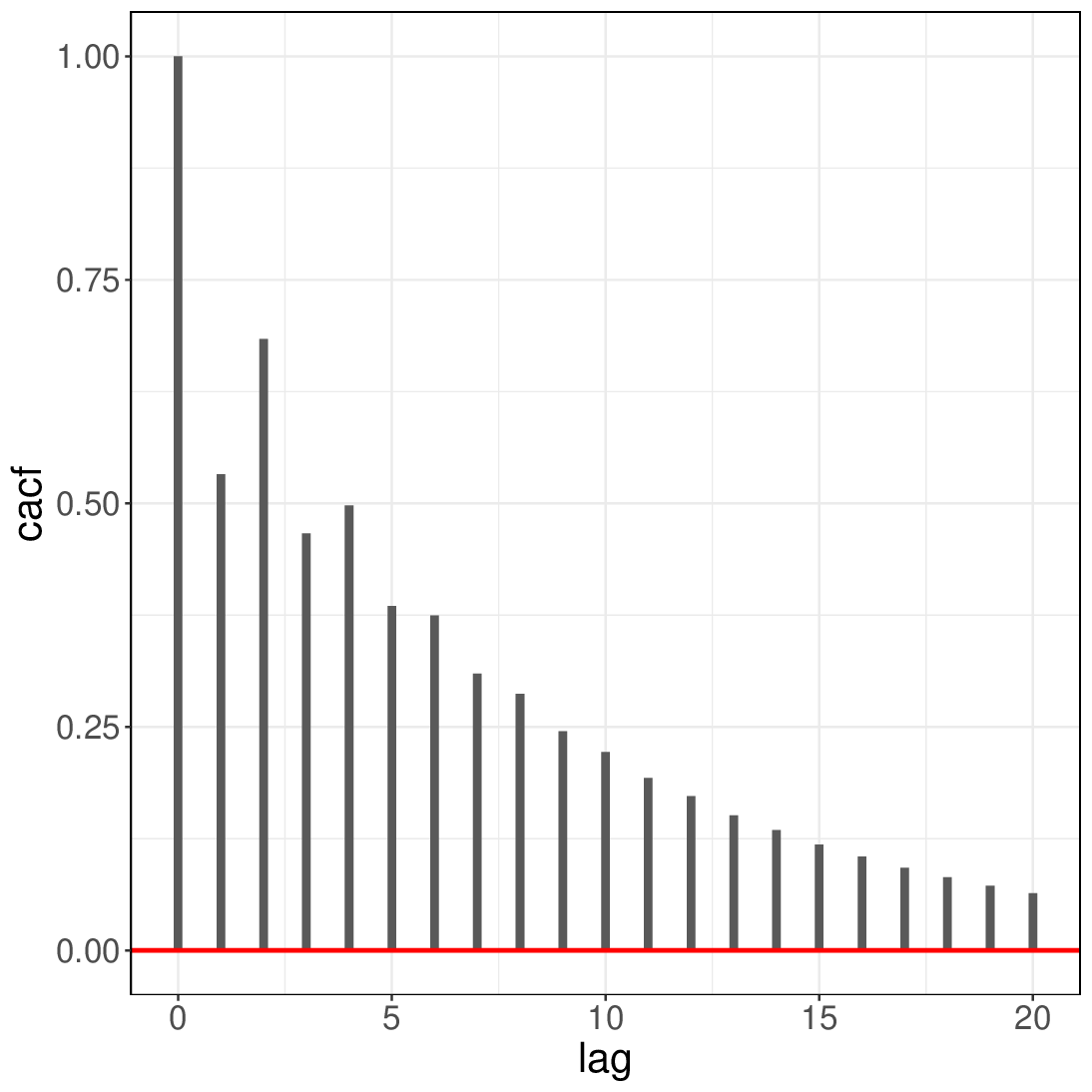}\\
\centering (a) $(q_1,q_2)=(1,1)$ 
 \end{minipage}
 \begin{minipage}{0.45\hsize}
  \centering
  \includegraphics[width=.9\textwidth,height=0.15\textheight]{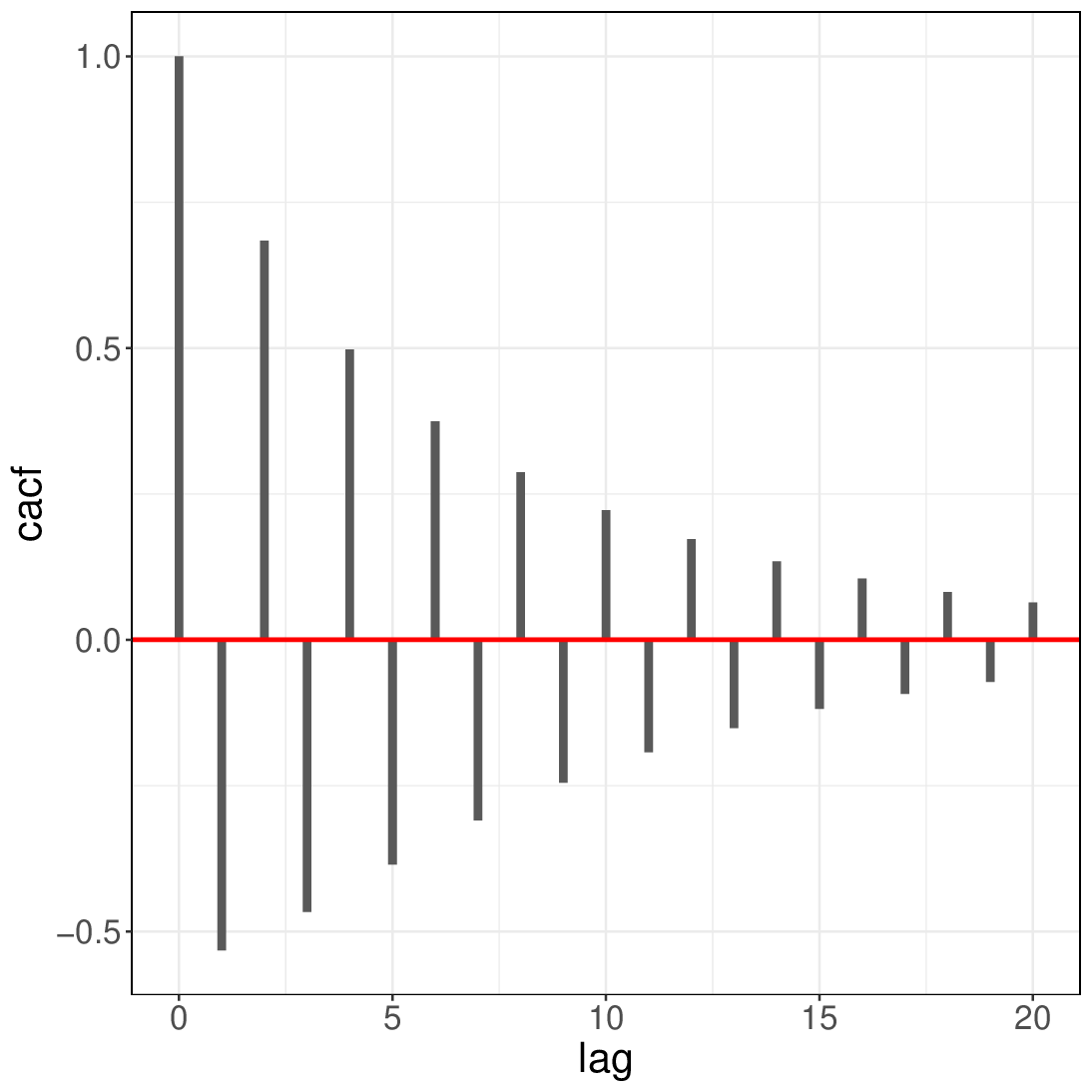}\\
\centering (b) $(q_1,q_2)=(-1,1)$ 
 \end{minipage}\\
 \begin{minipage}{0.45\hsize}
  \centering
  \includegraphics[width=.9\textwidth,height=0.15\textheight]{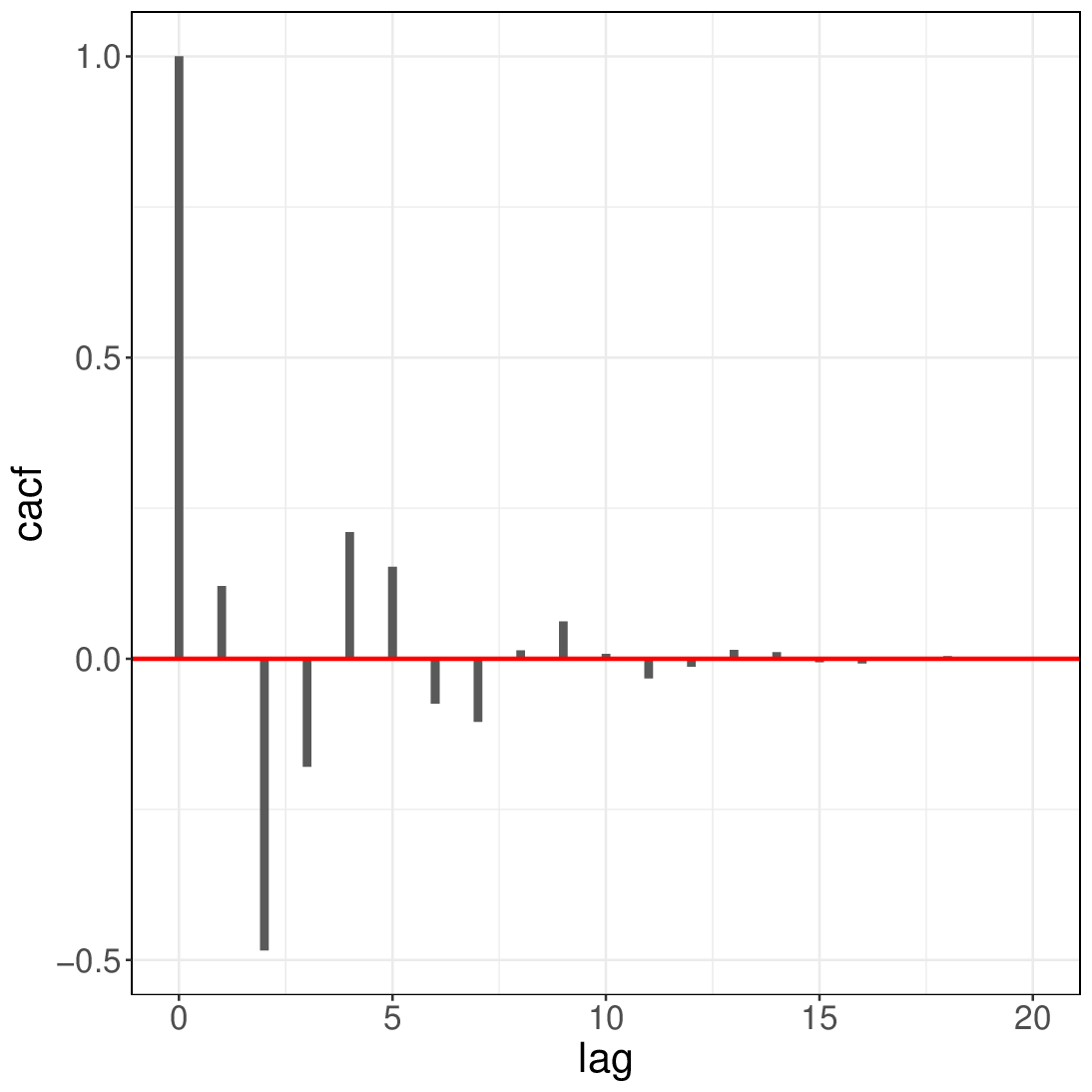}\\
\centering (c) $(q_1,q_2)=(1,-1)$ 
 \end{minipage}
 \begin{minipage}{0.45\hsize}
  \centering
  \includegraphics[width=.9\textwidth,height=0.15\textheight]{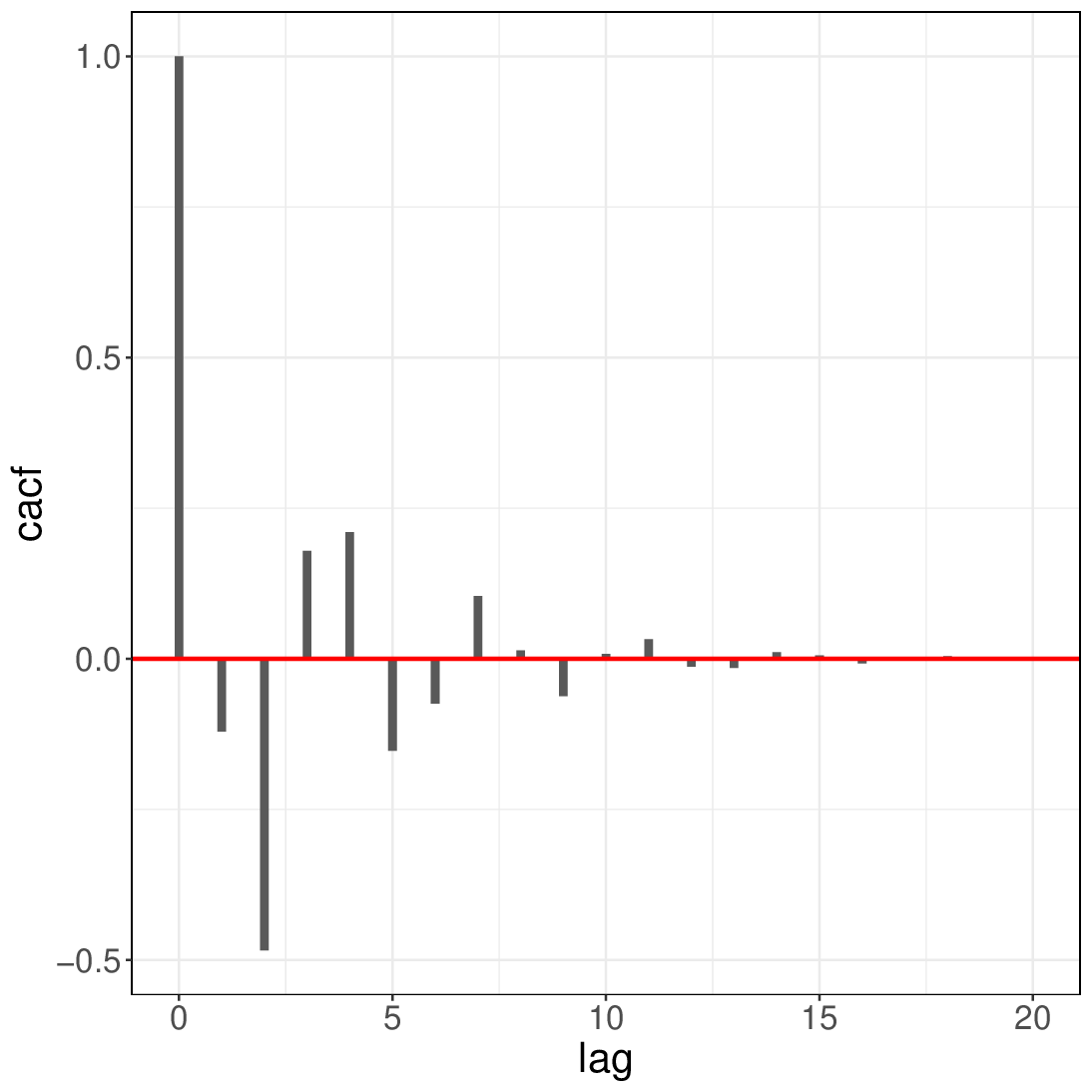}\\
\centering(d) $(q_1,q_2)=(-1,-1)$ 
 \end{minipage}
 \caption{Plots of the CACF of the MTD-AR(2) processes with parameters $a_1=0.3, a_2=0.7, \rho=0.9$, for panels (a)--(d), we set $(q_1,q_2)=(1,1)$, $(q_1,q_2)=(-1,1)$, $(q_1,q_2)=(1,-1)$, $(q_1,q_2)=(-1,-1)$, respectively.}\label{fig4}
\end{figure}

This section briefly illustrates a numerical example of the statistical properties provided in the previous sections. First, we plot the CACF of the MTD-AR(2) processes. The mixing weight parameter $\bm{a}$ is set as $(a_1, a_2)=(0.3, 0.7)$, and the binding density $g$ is chosen as a wrapped Cauchy distribution with concentration parameter $\rho=0.9$. The direction of the effects of the lagged variables is controlled by the parameter vector $\bm{q}=(q_1,q_2)$, which is set as $(q_1,q_2)\in \{(1,1), (-1,1), (1,-1), (-1,-1)\}$. Panels (a)--(d) in Figure~\ref{fig4} show the theoretical CACF given in (\ref{eq:CACF}) for four combinations of the signs in the parameters $(q_1, q_2)$. From panel (a), we can confirm that the positively autocorrelated time series is generated by setting positive signs of $q_i~i=1,2$. In contrast, from panel (b), we deduce that the negatively associated time-series characteristics are produced by setting negative $q_1$ and positive $q_2$. Panels (c) and (d) confirm the cyclical fluctuations when the parameter $q_2$ has a negative sign.

\begin{figure}[h]
 \begin{minipage}{0.45\hsize}
  \centering
  \includegraphics[width=.9\textwidth,height=0.15\textheight]{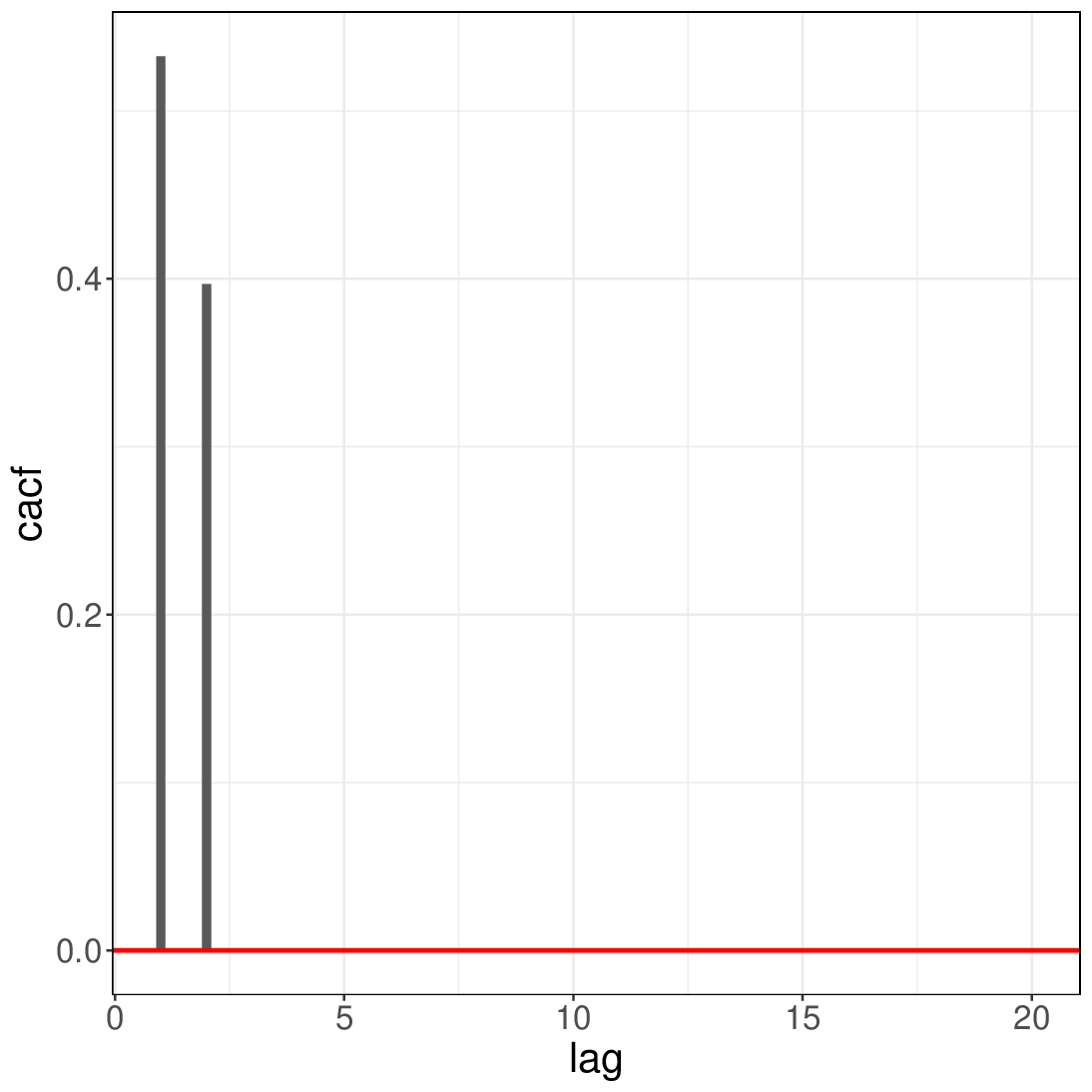}\\
\centering (a) $(q_1,q_2)=(1,1)$ 
 \end{minipage}
 \begin{minipage}{0.45\hsize}
  \centering
  \includegraphics[width=.9\textwidth, height=0.15\textheight]{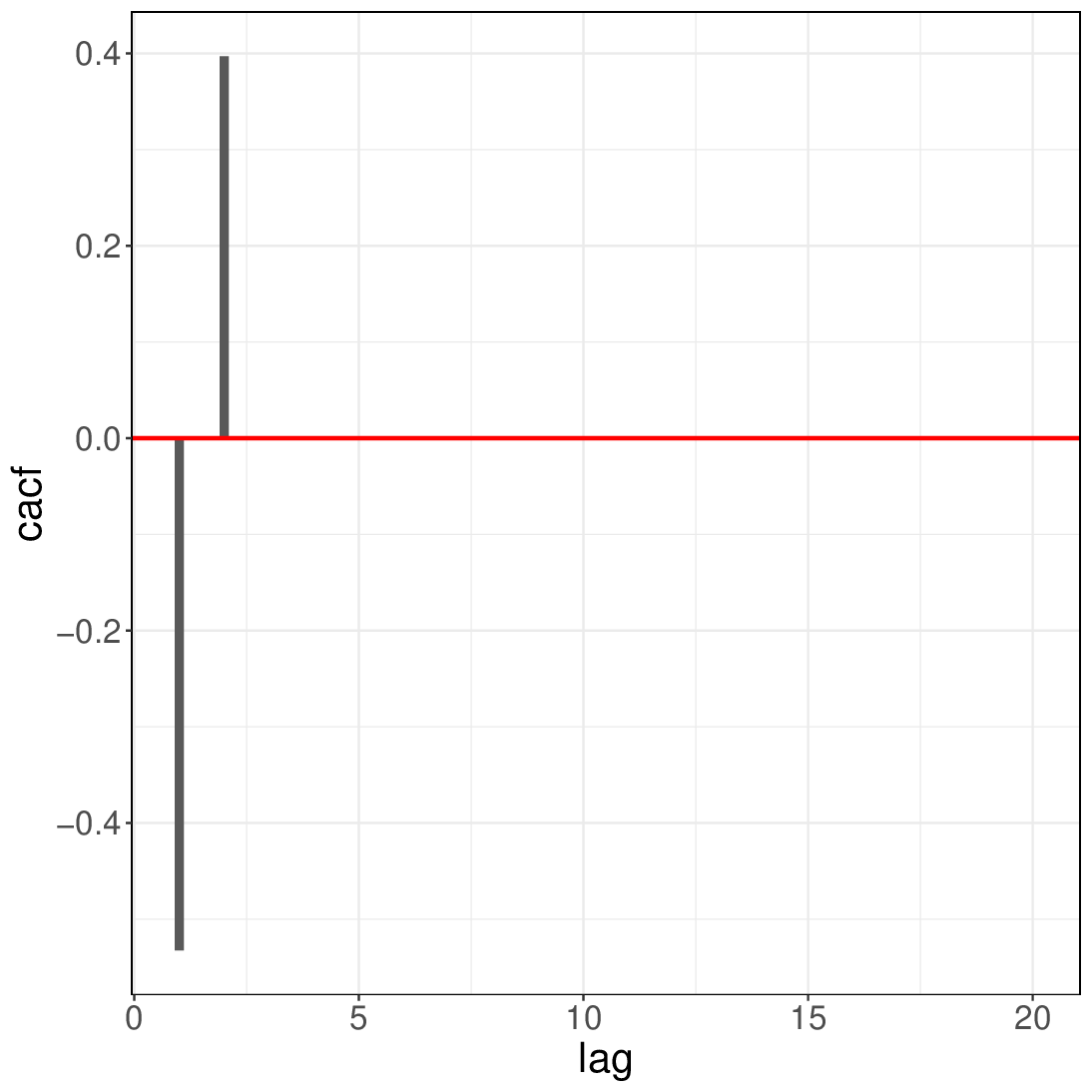}\\
\centering (b) $(q_1,q_2)=(-1,1)$ 
 \end{minipage}\\
 \begin{minipage}{0.45\hsize}
  \centering
  \includegraphics[width=.9\textwidth,height=0.15\textheight]{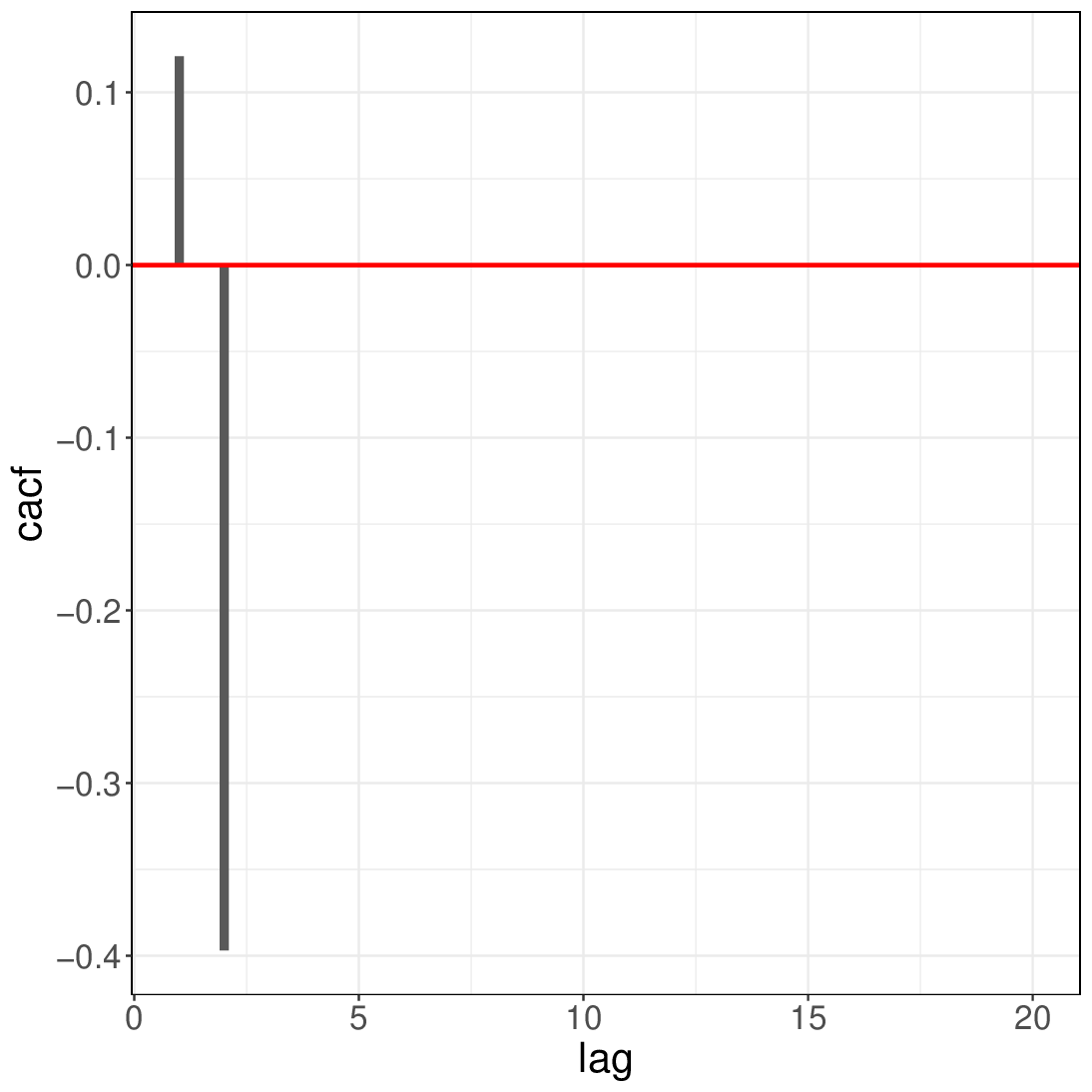}\\
\centering (c) $(q_1,q_2)=(1,-1)$ 
 \end{minipage}
 \begin{minipage}{0.45\hsize}
  \centering
  \includegraphics[width=.9\textwidth,height=0.15\textheight]{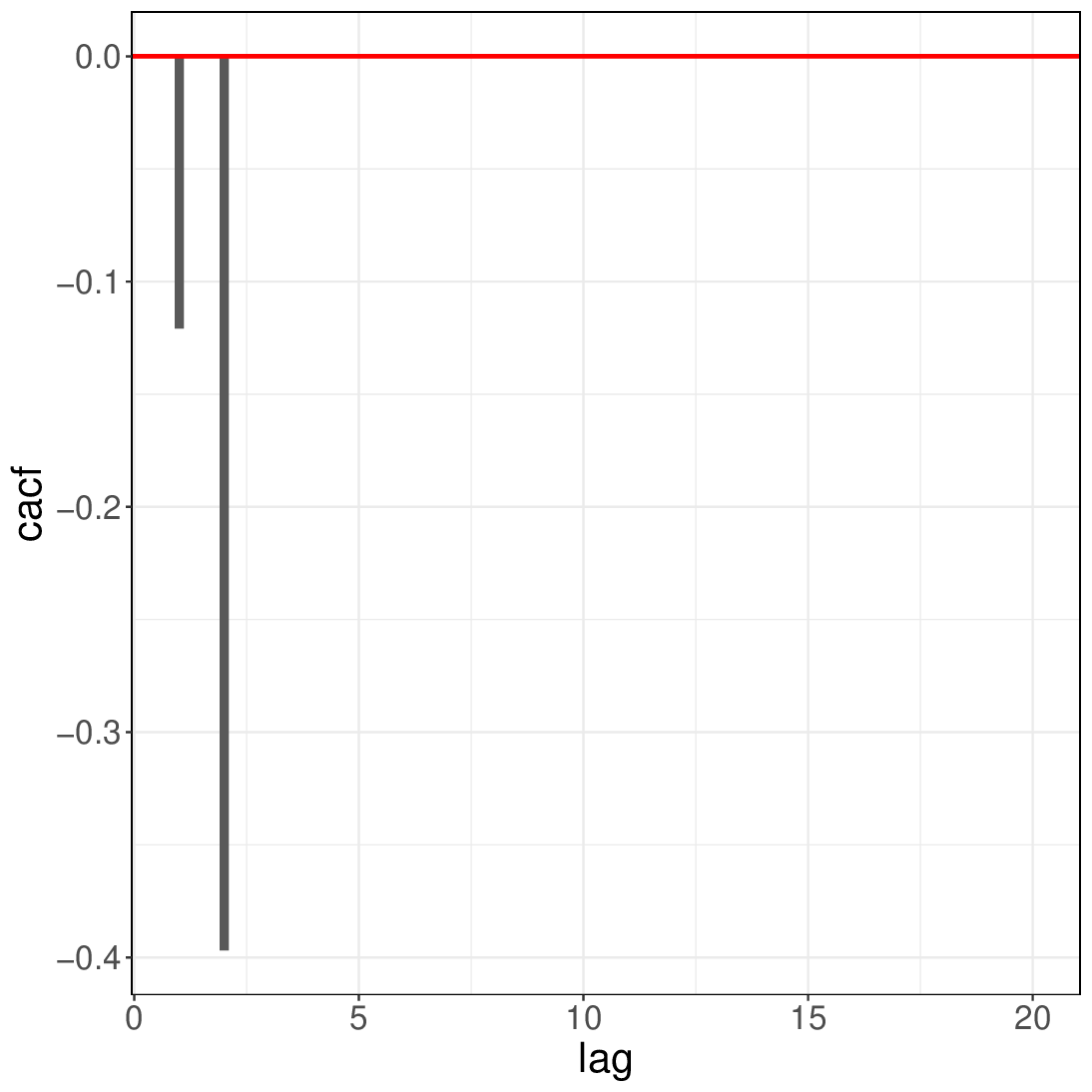}\\
\centering(d) $(q_1,q_2)=(-1,-1)$ 
 \end{minipage}
 \caption{Plots of the CPACF of the MTD-AR(2) processes with parameters $a_1=0.3, a_2=0.7, \rho=0.9$, for panels (a)-(d), we set $(q_1,q_2)=(1,1)$, $(q_1,q_2)=(-1,1)$, $(q_1,q_2)=(1,-1)$, $(q_1,q_2)=(-1,-1)$, respectively.}\label{fig5}
\end{figure}

Next, we investigate the CPACF of the MTD-AR(2) processes. Figure~\ref{fig5} shows the CPACF given by (\ref{eq:CPACF(det form)}) for the parameters used in Figure~\ref{fig4}. As it is a second-order Markov process, CPACFs with lags at greater than two vanish. The magnitude of the effect of the first two lag sequences is determined by the signs of the associated directions $\bm{q}$ and mixing weight $\bm{a}$ and the concentration parameters in the binding density $g$.

\begin{figure}[h]
  \centering
 \includegraphics[height=0.4\textheight, width=.9\textwidth]{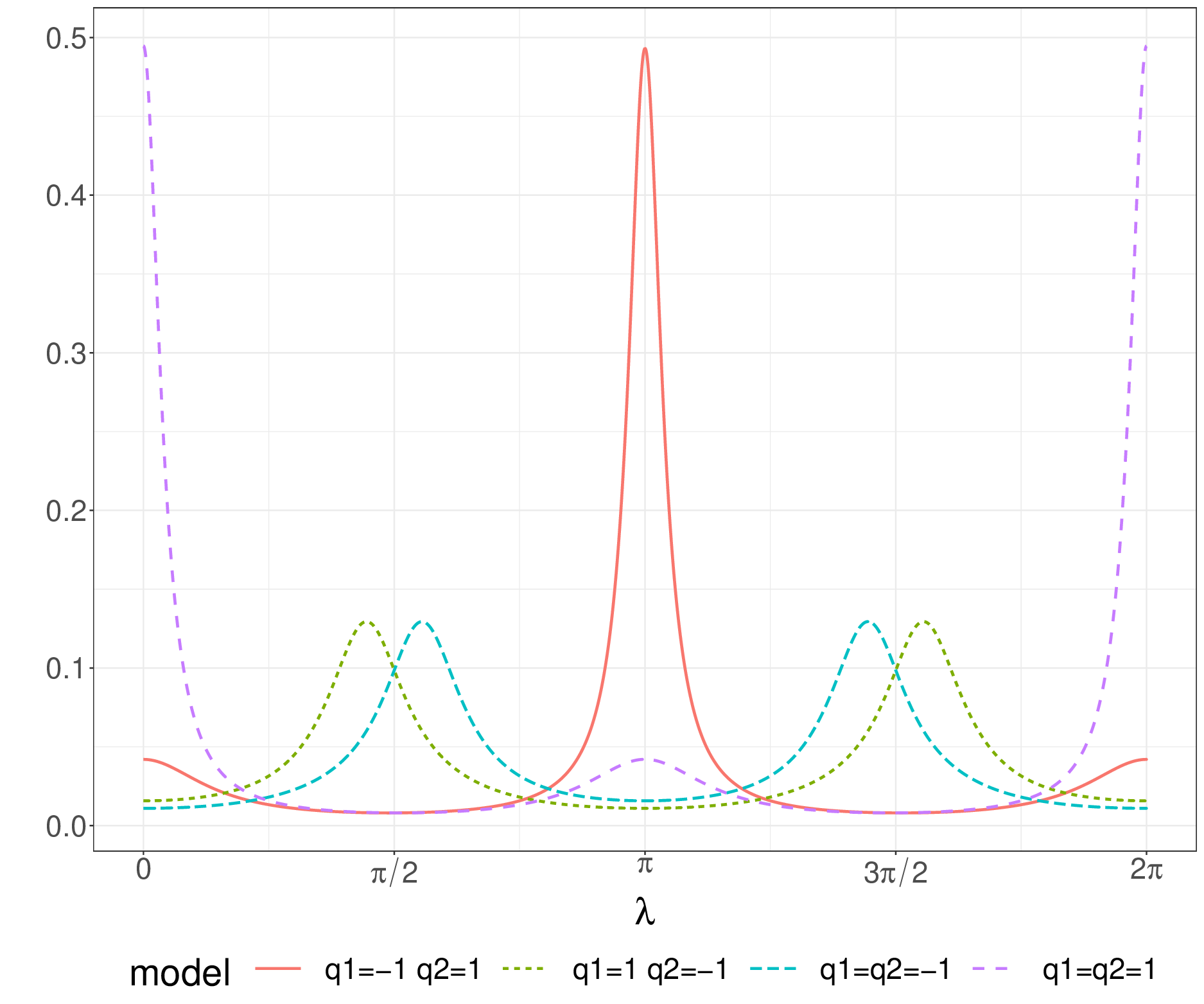}
 \caption{Plots of the spectral density function of MTD-AR(2) processes for $(q_1,q_2)=(1,1)$, $(q_1,q_2)=(-1,1)$, $(q_1,q_2)=(1,-1)$, $(q_1,q_2)=(-1,-1)$. The parameters $a_1, a_2$, and $\rho$ are the same as those used in Figure~\ref{fig4}.}
 \label{fig9}
\end{figure}

The corresponding spectral density functions are plotted in Figure~\ref{fig9}, where the parameters are the same as those used in the CACF plotted in Figure~\ref{fig4}. The spectral densities plotted here have a mode at some frequencies. These frequencies tell us the periodic patterns of time-series fluctuations examined from the CACF plots in Figure~\ref{fig4}.

The finite sample performance of the MLE is investigated using the parameters used in the numerical examples. The true model was MTD-AR(2) with a wrapped Cauchy density function as its binding density. The parameter vector was chosen as $\bm{\eta} = (a_1, \rho)= (0.3, 0.9)^T$. Recall that the mixing weight at lag 2 was $a_2=1-a_1=0.7$. The direction of the dependency parameter vector $q$ was set as $(1, 1)$, $(1, -1)$, $(-1, 1)$, and $(-1, -1)$. We drew a time series with length $n \in \{100, 250, 500\}$ from the MTD-AR(2) processes for the abovementioned parameters.

The finite sample performance of the MLE is investigated using the parameters used in the numerical examples. The true model was MTD-AR(2) with a wrapped Cauchy density function as its binding density. The parameter vector was chosen as $\bm{\eta} = (a_1, \rho)= (0.3, 0.9)^T$. Recall that the mixing weight at lag 2 was $a_2=1-a_1=0.7$. The direction of the dependency parameter vector $q$ was set as $(1, 1)$, $(1, -1)$, $(-1, 1)$, and $(-1, -1)$. We drew a time series with length $n \in \{100, 250, 500\}$ from the MTD-AR(2) processes for the parameters given above. 

First, we investigated the consistency and unbiased properties of the MLE for correctly specified cases, assuming that the autoregressive order of two with the true binding density $g$ was known in advance. The simulations were conducted with 1000 repetitions, and we calculated the average and the root mean squared error (RMSE) of the estimates. Table~\ref{tab1} summarizes the simulation results for the correctly specified case with known binding density $g$ and autoregressive order $p$. Table~\ref{tab1} shows that as $n$ increases, the RMSE decreases for all cases, indicating the consistency of the MLE. It was also confirmed that the estimation bias decreases as $n$ increases. We saw that the RMSEs of $\hat{a}_1$ with $\bm{q}=(1,1), (-1, 1)$ had larger values than those with $\bm{q}=(1,-1), (-1, -1)$.

\begin{table}[ht]
\centering
\caption{Simulation results for the maximum likelihood estimators for correctly specified case}. 
\label{tab1}
\begin{tabular}{lrrrrrrrr}
  \hline 
  & \multicolumn{4}{l}{$\hat{a}_1$} & \multicolumn{4}{l}{$\hat{\rho}$}\\
    $q$& $(1,1)$ & $(1,-1)$ & $(-1,1)$ & $(-1,-1)$ & $(1,1)$ & $(1,-1)$ & $(-1,1)$ & $(-1,-1)$ \\
    \hline
& \multicolumn{8}{l}{$n=100$}\\
Mean &  0.3071 & 0.3020 & 0.3057 & 0.3001 & 0.8991 & 0.8986 & 0.8993 & 0.8985\\ 
RMSE & 0.0767 & 0.0575 & 0.0747 & 0.0610 & 0.0147 & 0.0153 & 0.0147 & 0.0146\\ 
& \multicolumn{8}{l}{$n=250$}\\
Mean & 0.3051 & 0.3015 & 0.3023 & 0.3002 & 0.8997 & 0.8998 & 0.8999 & 0.8999 \\ 
RMSE & 0.0489 & 0.0391 & 0.0492 & 0.0376 & 0.0089 & 0.0088 & 0.0091 & 0.0088\\ 
& \multicolumn{8}{l}{$n=500$}\\
 Mean & 0.3023 & 0.3004 & 0.3022 & 0.3014 & 0.8997 & 0.8996 & 0.9005 & 0.8999 \\ 
 RMSE & 0.0341 & 0.0252 & 0.0352 & 0.0260 & 0.0062 & 0.0064 & 0.0062 & 0.0065\\ 
   \hline
\end{tabular}
\end{table}

Second, we investigated a misspecified case in which the underlying binding density function was misspecified from a true binding density. We chose a true binding density as a wrapped Cauchy distribution and von Mises and Jones--Pewsey distributions as misspecified cases. In practice, we do not have enough information about the true density, so a flexible distribution containing many important distributions should be applied.  Hence, the Jones--Pewsey distribution is a natural candidate for the underlying binding density function. For more details about Jones--Pewsey distributions, we refer to \cite{jones2005family}. The scale parameter $\kappa$ of the von Mises distribution and the scale and shape parameters $\kappa$ and $\psi$ of the Jones--Pewsey distribution are transformed into first-order cosine moment via $\rho_{VM}= I_1(\kappa)/ I_0(\kappa)$ and 
\begin{align*}
\rho_{JP}= \frac{|\psi|}{1+\psi} \frac{P^1_{1/\psi}(\cosh (\kappa \psi))}
{P_{1/\psi}(\cosh (\kappa \psi))},
\end{align*}
respectively. Here, $I_{\nu}(\kappa)$ is the modified Bessel functions of the first kind of order $\nu$ and $P_{s}^m(x)$ is the associate Legendre polynomial function with order $m$ and degree $s$.
Table~\ref{tab2} summarizes the simulation results of the MLE when the true binding density $g$ was the wrapped Cauchy distribution while we applied the von Mises and Jones--Pewsey distributions as the binding density in log-likelihood maximization. The parameter values were the same as those used in Table~\ref{tab1}. According to Table~\ref{tab2}, the MLE with the Jones--Pewsey distributions has smaller RMSEs compared to those with von Mises distribution. However, the RMSEs have larger values compared with those in Table~\ref{tab1} for both distributions. This indicates that the Jones--Pewsey models could not achieve a lower bound of MLE even though this model contains the wrapped Cauchy distribution. The reason behind this can be explained twofold. First, the Jones--Pewsey distribution is a three-parameter distribution, whose shape parameter $\psi$ controls the probability around the mode and anti-mode. It can be observed numerically and theoretically that the variance of the estimated concentration parameter $\hat{\kappa}$ of the Jones--Pewsey distribution is larger than that of the $\hat{\kappa}$ and $\hat{\rho}$ for von Mises and wrapped Cauchy distributions in the case with fixed $\psi=0$ and $\psi=-1$, respectively. This phenomenon is more significant when $\rho$ approaches one. Hence, the three-parameter Jones--Pewsey distribution is less effective than the wrapped Cauchy distribution in the case with  correctly specified settings, whereas it is more effective compared to the von Mises distribution in the case with  completely misspecified settings. Second, when data shows a concentrated distributional pattern, the correlation coefficient between $\hat{\kappa}$ and $\hat{\psi}$ approaches one for Jones--Pewsey distribution. This may cause the diverging variance of the $\hat{\kappa}$ for the Jones--Pewsey distribution for some given data. Actually, several simulated results show that the estimated parameters $\hat{\kappa}$ have much larger values and yield larger RMSE.
Hence, we saw that the MLE for the misspecified case did not provide an efficient estimation for the model parameters, whereas this method can provide consistent estimates for unknown parameters $\bm{\eta}=(a_1, \rho)$.

For almost all cases, the RMSEs of $\hat{\rho}_{VM}$ and $\hat{\rho}_{JP}$ for $\bm{q} =(1,1)$ and $(-1,1)$ have smaller values than those with $\bm{q} =(1,-1)$ and $(-1,-1)$. In contrast, the RMSEs of $\hat{a}_1$ for $\bm{q} =(1,1)$ and $(-1, 1)$ have larger values than those with $\bm{q} =(1,-1)$ and $(-1,-1)$.

\begin{table}[ht]
\centering
\caption{Simulation results for the maximum likelihood estimates under the misspecified case.}
\label{tab2}
\begin{tabular}{lrrrrrrrr}
\hline 
  & \multicolumn{4}{l}{$\hat{a}_1$} & \multicolumn{4}{l}{$\hat{\rho}_{VM}$}\\
    $q$& $(1,1)$ & $(1,-1)$ & $(-1,1)$ & $(-1,-1)$ & $(1,1)$ & $(1,-1)$ & $(-1,1)$ & $(-1,-1)$ \\
    \hline
& \multicolumn{8}{l}{$n=100$}\\
Mean  &  0.3031 & 0.3036 & 0.3020 & 0.3005 & 0.9003 & 0.9086 & 0.8998 & 0.9087 \\ 
RMSE &  0.0981 & 0.0615 & 0.0995 & 0.0656 & 0.0300 & 0.0295 & 0.0302 & 0.0297  \\ 
& \multicolumn{8}{l}{$n=250$}\\
Mean  &  0.2954 & 0.3028 & 0.2925 & 0.3017 & 0.9016 & 0.9126 & 0.9007 & 0.9128  \\ 
RMSE& 0.0489 & 0.0391 & 0.0492 & 0.0376 & 0.0195 & 0.0224 & 0.0194 & 0.0216\\ 
& \multicolumn{8}{l}{$n=500$}\\
Mean  & 0.2916 & 0.3025 & 0.2908 & 0.3039 & 0.9002 & 0.9119 & 0.9016 & 0.9120 \\ 
RMSE & 0.0341 & 0.0252 & 0.0352 & 0.0260 & 0.0136 & 0.0180 & 0.0138 & 0.0181 \\ 
   \hline
   & \multicolumn{4}{l}{$\hat{a}_1$} & \multicolumn{4}{l}{$\hat{\rho}_{JP}$}\\
   & \multicolumn{8}{l}{$n=100$}\\
Mean  & 0.3069 & 0.3019 & 0.3055 & 0.3001 & 0.9007 & 0.8976 & 0.8993 & 0.8999 \\ 
RMSE &  0.0770 & 0.0576 & 0.0746 & 0.0611 & 0.0296 & 0.0300 & 0.0291 & 0.0294 \\ 
& \multicolumn{8}{l}{$n=250$}\\
Mean  & 0.3050 & 0.3016 & 0.3022 & 0.3003 & 0.9010 & 0.9010 & 0.9000 & 0.9009 \\ 
RMSE &  0.0489 & 0.0391 & 0.0493 & 0.0377 & 0.0185 & 0.0186 & 0.0176 & 0.0183 \\ 
& \multicolumn{8}{l}{$n=500$}\\
 Mean  & 0.3022 & 0.3004 & 0.3022 & 0.3013 & 0.8995 & 0.8996 & 0.9011 & 0.8997 \\ 
RMSE &  0.0341 & 0.0252 & 0.0351 & 0.0260 & 0.0125 & 0.0135 & 0.0128 & 0.0139 \\ 
\hline
\end{tabular}
\end{table}

Finally, we investigated the estimation results of the unknown autoregressive order. We employed the AIC and BIC. The true model has an autoregressive order $p=2$ with parameters $\bm{\eta} = (a_1, \rho)= (0.7, 0.9)^T$ with wrapped Cauchy binding densities. The sample size was $n= \{ 100, 250, 500\}$, and Monte Carlo simulations were performed with 1000 replications. Table~\ref{tab3} summarizes the number of selected models by AIC and BIC with a grid search of the unknown autoregressive orders from 1 to 5. In the case with $q_1=-1$, no models were selected lags at 1 for all cases, whereas for $q_1=1$, only a few were selected with $\hat{p}=1$ for AIC and BIC. We observe that the BIC is the consistent estimator of $\hat{p}$ as the ratio of the corrected models selected increases as $n$ increases, whereas the AIC works worse than the BIC.

\begin{table}[ht]
\centering
\caption{The number of selected models by AIC and BIC among 1000 times replications. The true model has $p=2$ and wrapped Cauchy binding densities with parameters $\bm{\eta} = (a_1, \rho)= (0.7, 0.9)^T$.}
\label{tab3}
\begin{tabular}{l rrrrr rrrrr}
\hline 
      & \multicolumn{5}{l}{AIC} & \multicolumn{5}{l}{BIC}   \\  
      $p$  &  1 & 2 & 3 & 4 & 5 & 1 & 2 & 3 & 4 & 5  \\
      \hline 
      & \multicolumn{10}{l}{$\bm{q}=(1,1)$}\\
 $n=100$ &   7  & 145  & 359  & 229 &  260  & 31 &  336  & 348  &156  & 129 \\
 $n=250$ &    0  & 207  & 304  & 230  & 259  &  2  & 564  & 284   & 89   & 61 \\
 $n=500$ &    0  & 258  & 287 & 212  & 243 &    0  & 669 &  212   & 77   & 42 \\
   \hline
         & \multicolumn{10}{l}{$\bm{q}=(1,-1)$}\\
 $n=100$ &   13  & 150 &  338  & 251  & 248 &  43  & 327  & 339 & 173 &  118 \\
 $n=250$ &     0   & 241 &  289 &  228  & 242  &  1  & 558  & 254 & 113  & 74  \\
 $n=500$ &   0  & 274 & 323 & 179 &   224 &   0  &700 & 221  & 43   & 36  \\
   \hline
         & \multicolumn{10}{l}{$\bm{q}=(-1,1)$}\\
 $n=100$ &    0 &  237  & 372 & 181  & 210 &   0  & 498  & 342  & 97 &   63\\
 $n=250$ &   0  & 256  & 322  & 227  & 195  &  0  & 651 & 248  & 74   & 27\\
 $n=500$ &    0  & 288 & 328  & 173 & 211   & 0 & 708 & 230   & 45   & 17\\
   \hline
         & \multicolumn{10}{l}{$\bm{q}=(-1,-1)$}\\
 $n=100$ &   0 & 215  & 380 & 211 & 194 &   0  & 495 &  328  & 126 &   51\\
 $n=250$ &  0  & 265  & 336  & 192  & 207 &   0  & 667 & 235  & 60   & 38\\
 $n=500$ &   0  & 313 &  314  & 185 & 188  &  0 & 730 & 212 &  48   & 10\\
   \hline
\end{tabular}
\end{table}

\section{Data Analysis}\label{sec:6}

\begin{figure}[H]
  \centering
 \includegraphics[height=5cm, width=.8\textwidth]{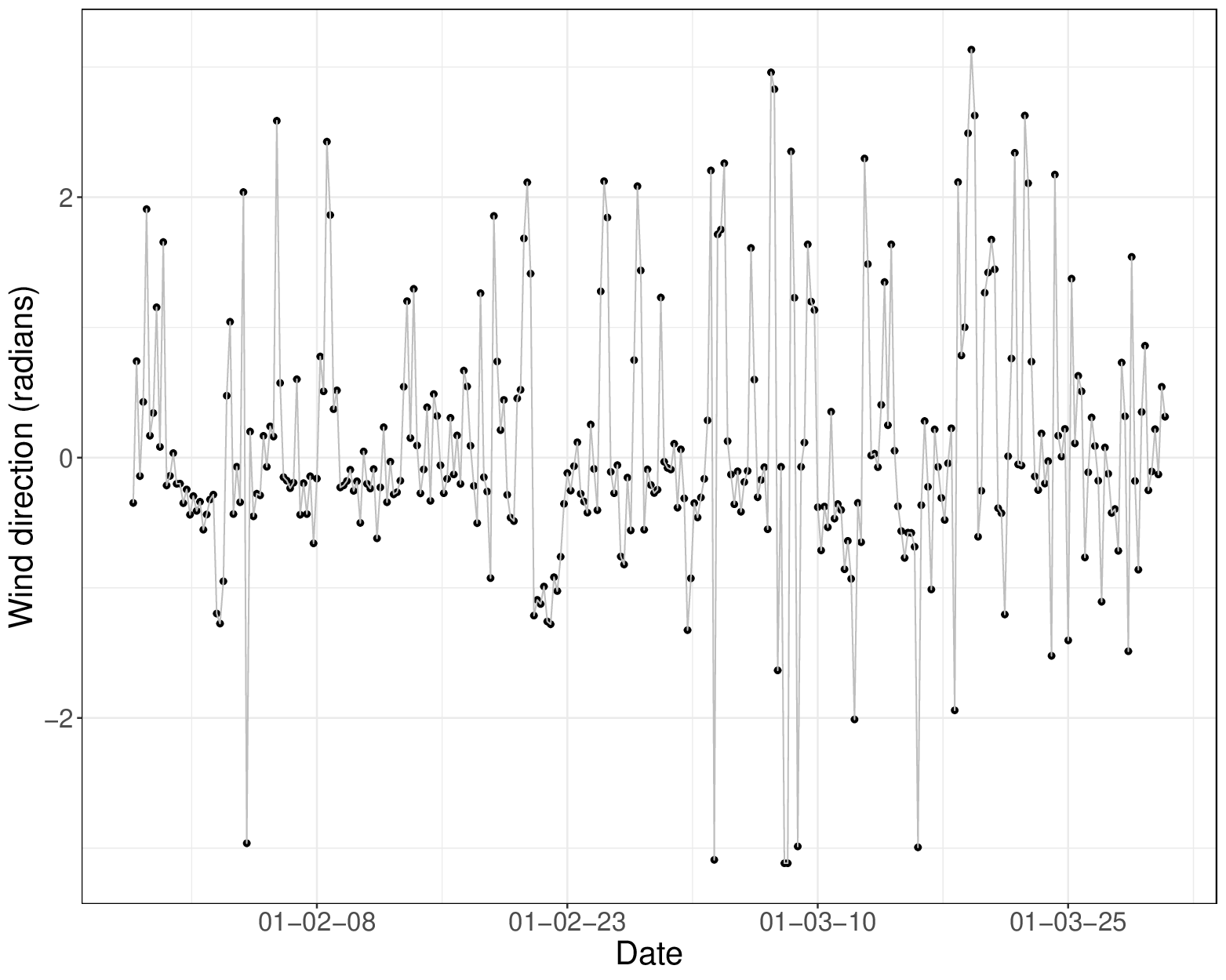}\\
 \caption{Time-series plots for the wind direction at Col de la Roa in Italy.}
 \label{figA1}
\end{figure}

\begin{figure}[H]
  \centering
 \includegraphics[height=5cm, width=.8\textwidth]{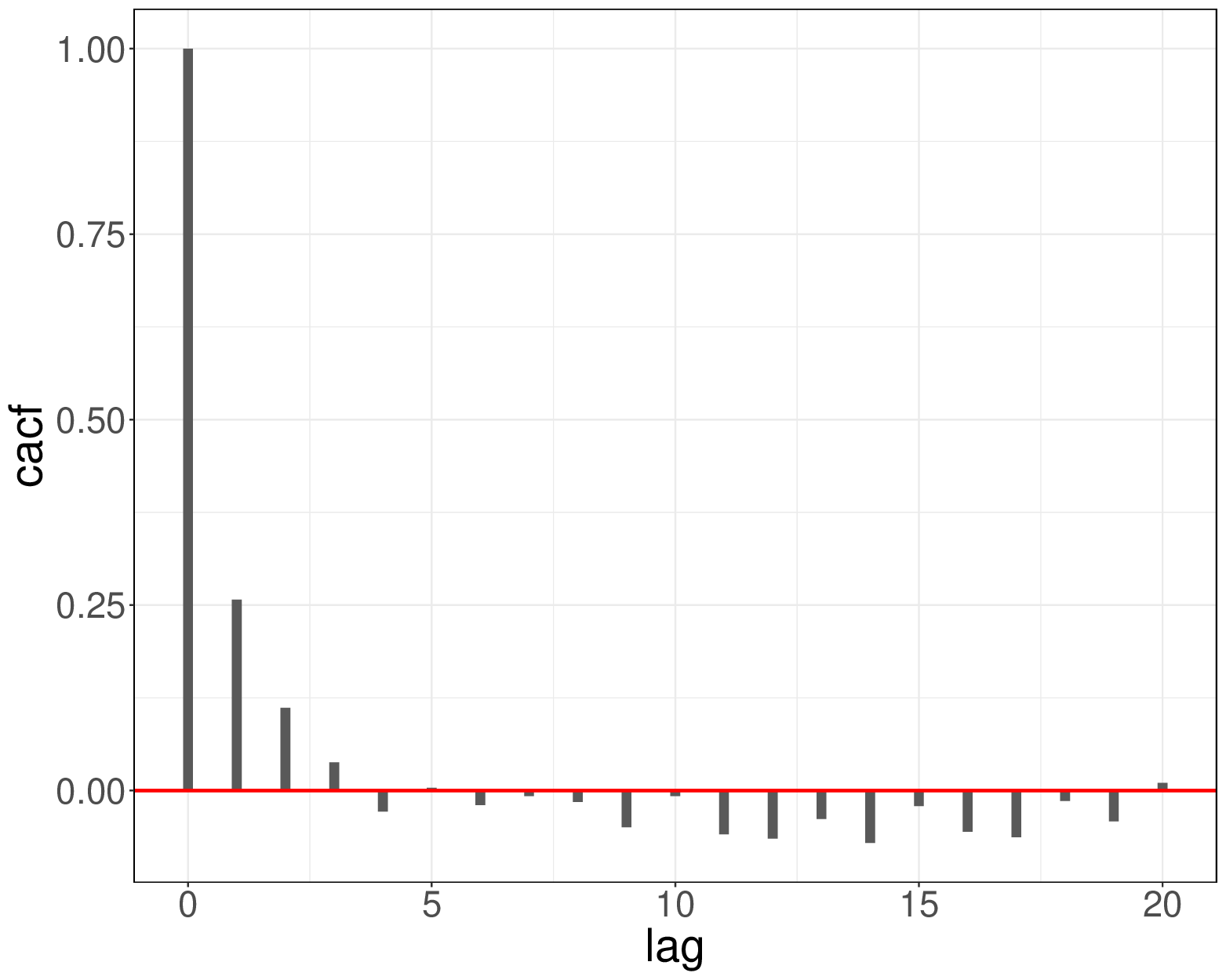}\\
 \caption{Sample CACF for the wind direction at Col de la Roa in Italy.}
 \label{figA2}
\end{figure}

\begin{figure}[H]
  \centering
 \includegraphics[height=5cm, width=.8\textwidth]{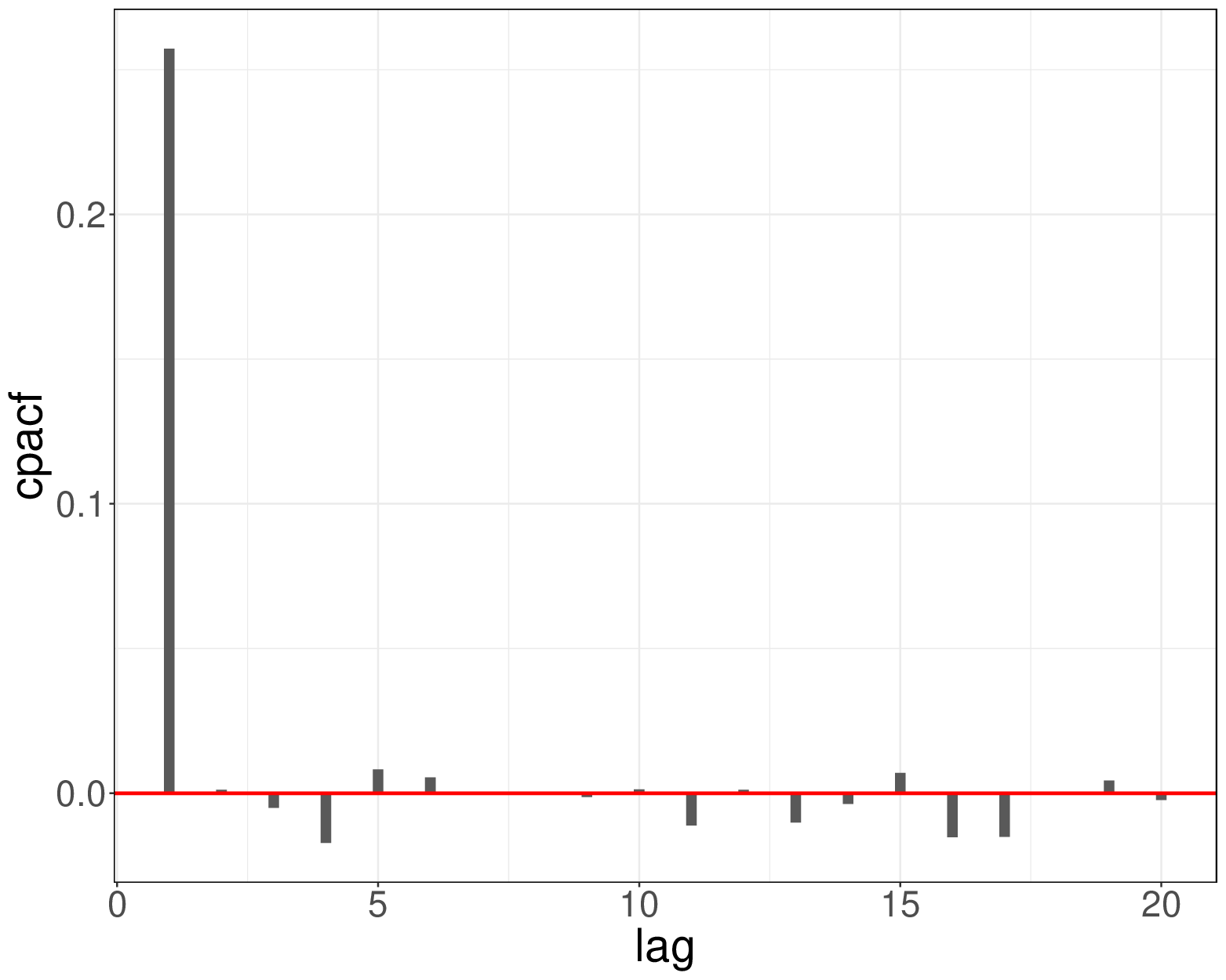}\\
 \caption{Sample CPACF for the wind direction at Col de la Roa in Italy.}
 \label{figA3}
\end{figure}

We fitted the MTD-AR($p$) models for wind direction data at Col de la Roa in Italy, investigated by \cite{agostinelli2007robust}. The data consisted of 310 measurements of wind directions from January 29 to March 31, 2001, and the data were collected every 15 minutes from 3.00 am to 4.00 am. As confirmed from a time-series plot shown in Figure~\ref{figA1}, moderately positive autocorrelation existed, which was verified from the sample CACF plotted in Figure~\ref{figA2}. From Figure~\ref{figA2}, we also observed that the negative circular autocorrelation existed at larger lags around 9 through 19. Figure~\ref{figA3} shows the sample CPACF of the observed time series. According to this figure, we see that a positive partial correlation at lag 1 and a weakly negative partial correlation at lag 4 exist, making the observed time series moderately periodic patterns.  

\begin{figure}[H]
  \centering
 \includegraphics[height=5cm, width=.8\textwidth]{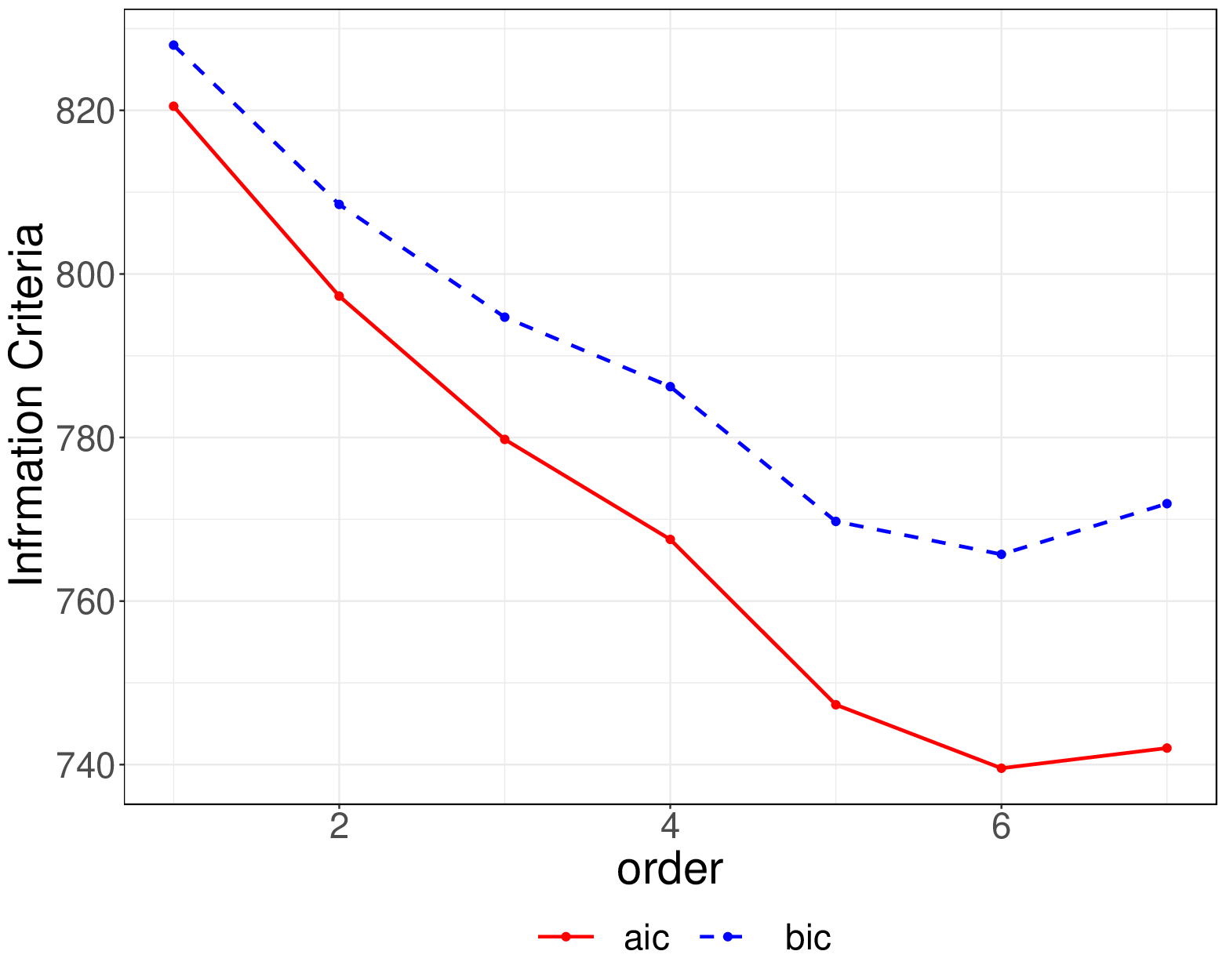}\\
 \caption{Plots of the AIC and BIC for fitted MTD-AR($p$) models with $p=1,\ldots, 7$.}
 \label{figA4}
\end{figure}

The unknown autoregressive order $p$ was determined by AIC and BIC, and we chose the wrapped Cauchy distribution as the binding density $g$. Figure~\ref{figA4} plots the values of AIC and BIC against maximum lags at seven. According to the figure, the lag order was selected at six from both the AIC and BIC.

Table~\ref{tabA5} summarized the estimated parameters with mixture weights, the concentration of the binding density $g$, and the signs of the directions of the dependency at lags from 1 through 7. The estimated $\hat{q}_i$s were denoted with signs in the parenthesis, where $(+)$ and $(-)$ indicate $\hat{q}_i=+1$ and $\hat{q}_i=-1$, respectively. Recall that the MTD-AR($p$) models with lag order $p=6$ were selected as the best model among lags from 1 to 7.

\begin{center}
\begin{table}
\caption{Estimated mixture weights together with the directions of the dependency at lags from 1 through 7.}
\label{tabA5}
\begin{tabular}{l ccccccc c}
\hline
$p$ & $\hat{a}_1$ & $\hat{a}_2$ & $\hat{a}_3$ & $\hat{a}_4$ & $\hat{a}_5$ & $\hat{a}_6$ & $\hat{a}_7$  & $\hat{\rho}$\\ 
\hline
1 & 1.000 &  &  &  &  &  &   & 0.642\\ 
2 & $(+)$0.744 & $(+)$0.256 &  &  &  &  & & 0.689\\ 
3 & $(+)$0.654 & $(+)$0.186& $(-)$0.160 &  &  &  & & 0.721 \\ 
4 & $(+)$0.599 & $(+)$0.118 & $(+)$0.119 & $(+)$0.164 &  &  & &0.736\\ 
5 & $(+)$0.545 & $(+)$0.092 & $(+)$0.087 & $(-)$0.085 & $(+)$0.191 &  &  & 0.760\\ 
\textbf{6} & $(+)$0.514 & $(+)$0.086 & $(+)$0.072 & $(-)$0.085 & $(+)$0.160 & $(+)$0.083 & & 0.755\\ 
7 & $(+)$0.508 & $(+)$0.082 & $(+)$0.069 & $(+)$0.086 & $(+)$0.153 & $(+)$0.069 & $(+)$0.032& 0.773\\ 
\hline
\end{tabular}
\end{table}
\end{center}

\begin{figure}[H]
  \centering
 \includegraphics[height=6cm, width=.8\textwidth]{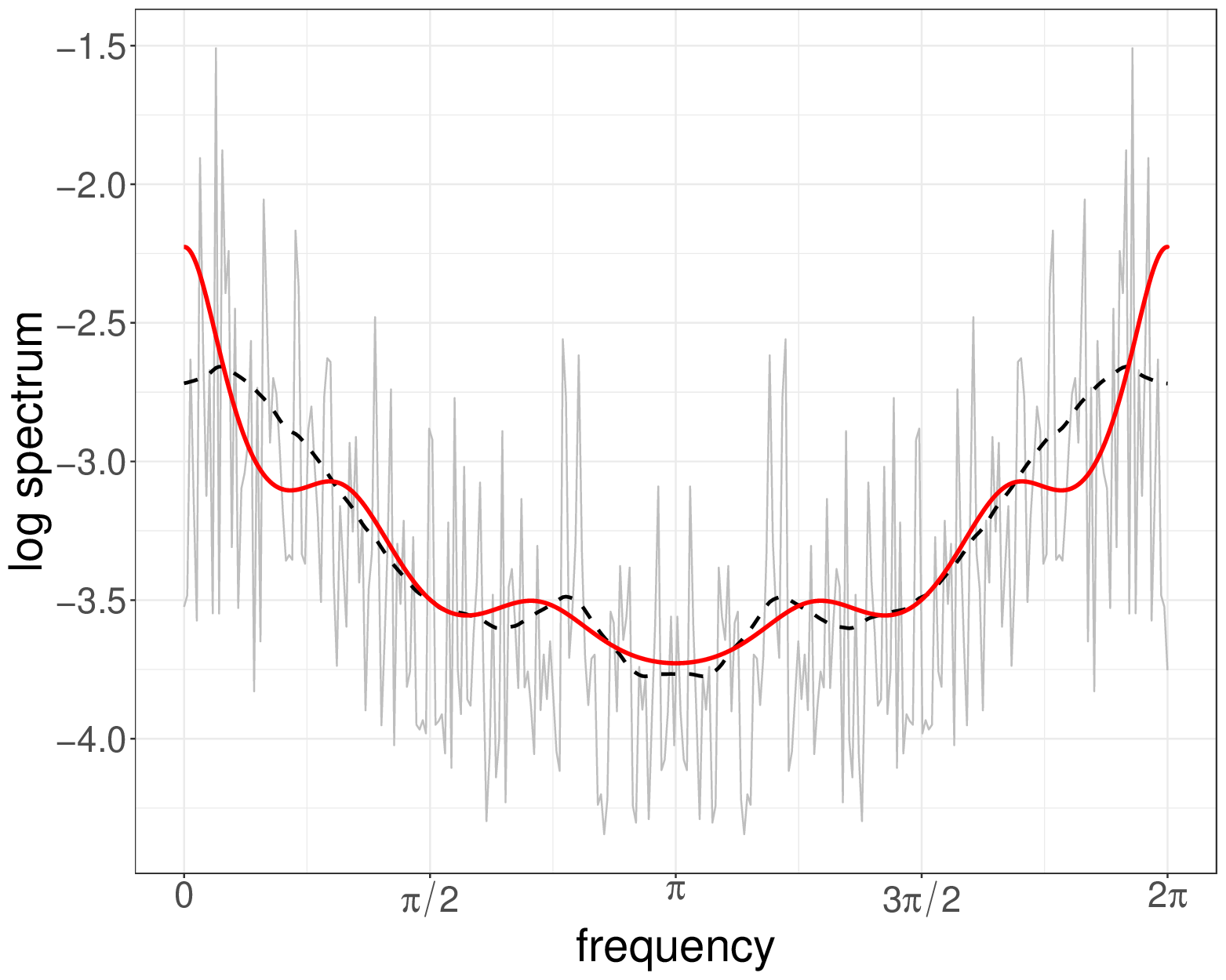}\\
 \caption{Plots of the circular periodogram, its smoothed estimates (dashed line), and the spectral density function of the fitted MTD-AR(6) model (solid line).} 
 \label{figA6}
\end{figure}

The log periodogram for the circular time series, together with its smoothed estimates and log spectral density function for the fitted MTD-AR(6), are illustrated in Figure~\ref{figA6}. The fitted MTD-AR(6) spectral density function performs well in capturing observed spectra, which can be confirmed by comparing the log-smoothed periodogram.

\section{Conclusion}\label{sec:7}
This study proposed a higher-order Markov model on the circle by incorporating a mixture of transition density modeling. Statistical properties, including first- and second-order stationarity together with CACF and CPACF, were derived. In addition, the spectral density function of the circular time-series processes was obtained by assuming that the location parameter of the binding density was fixed at $\mu=0$. For the parameter estimation, the MLE was introduced, and its asymptotic properties were provided. These theoretical results were investigated by a numerical illustration using the MTD-AR(2) process and Monte Carlo simulations were conducted to verify the finite sample performance of the MLEs. The limitations of the proposed models include that the explicit expressions for the spectral density function are for the case with $\mu=0$, whereas for the general case with $\mu\ne 0$, the CACF, together with the spectral density function, have complicated forms. These problems may inspire future research. In addition, non-parametric spectral density function estimators are worth investigating.


\section*{Acknowledgments} 
The authors are grateful to the editor and two anonymous referees for helpful
suggestions that led to the improvement of the paper.
Hiroaki Ogata was supported in part by JSPS KAKENHI Grant Numbers 18K11193 and 24K14858. Takayuki Shiohama was supported in part by JSPS KAKENHI Grant Number 22K11944 and Nanzan University Pache Research I-A-2 for the 2024 academic year.

\bibliographystyle{chicago}
\bibliography{MTDarxiv}

\appendix

\section{Appendix}\label{sec:A}
In this appendix, we provide all proofs of the theorems and lemmas stated in Sections~\ref{sec:3} and \ref{sec:4}. We give the following lemmas. 
\begin{lem}
	\label{lemma 1}
	Let $g$ be an arbitrary circular density satisfying Assumption 1. Then,
	\begin{align*}
	\int_{\Pi} 
	\left(
	\begin{array}{c}
	\cos m\theta_j \\
	\sin m\theta_j
	\end{array}
	\right)
	g(\theta_j-q_i \theta_{j-i}) d\theta_j
	=D_m
    Q_i
	\left(
	\begin{array}{c}
	\cos m\theta_{j-i} \\
	\sin m\theta_{j-i}
	\end{array}
	\right),
	\end{align*}
	where $\Pi = (-\pi,\pi]$, 
	\begin{align*}
	D_m
	=\rho_m 
	\left(
	\begin{array}{cc}
	\cos \mu_m & -\sin \mu_m \\
	\sin \mu_m & \cos \mu_m
	\end{array}
	\right), \quad
	Q_i =
\left(
\begin{array}{cc}
1 & 0 \\
0 & q_i
\end{array}
\right),
	\end{align*}
	and $ q_i \in \{-1,1\} $ is a constant, and $ \rho_m$ and $\mu_m $ are the resultant length and direction of $m$th order trigonometric moment of the circular density function $ g $.  
\end{lem}
\textbf{Proof} Direct calculation yields the following result:
\begin{align*}
&\int_{\Pi} 
\left(
\begin{array}{c}
\cos m\theta_j \\
\sin m\theta_j
\end{array}
\right)
g(\theta_j-q_i \theta_{j-i}) d\theta_j\\
=&
\int_{\Pi} 
\left(
\begin{array}{c}
\cos m(s+q_i \theta_{j-i}) \\
\sin m(s+q_i \theta_{j-i})
\end{array}
\right)
g(s) ds \\
=&
\int_{\Pi} 
\left(
\begin{array}{c}
\cos ms\cos(q_i m\theta_{j-i}) - \sin ms\sin(q_i m\theta_{j-i})  \\
\sin ms\cos(q_i m\theta_{j-i}) + \cos ms\sin(q_i m\theta_{j-i})
\end{array}
\right)
g(s) ds \\
=&
\int_{\Pi} 
\left(
\begin{array}{cc}
\cos ms & -\sin ms \\
\sin ms & \cos ms
\end{array}
\right)
g(s) ds \ 
\left(
\begin{array}{c}
\cos q_i m\theta_{j-i} \\
\sin q_i m\theta_{j-i}
\end{array}
\right)\\
=&
D_m Q_i
\left(
\begin{array}{c}
\cos m\theta_{j-i} \\
\sin m\theta_{j-i}
\end{array}
\right),
\end{align*}
where we use a transformation of variable in the second equation with $s=\theta_j -q_i\theta_{j-i}$ and $ds=d\theta_j$.\hfill $\square$

\begin{lem}
	\label{lemma 2}
	Let $g$ be an arbitrary circular density satisfying Assumption 1. Then,
	\begin{align*}
	\int_{\Pi} 
	\left(
	\begin{array}{rr}
	\cos m\theta_j & \sin m\theta_j \\
	\sin m\theta_j & -\cos m\theta_j
	\end{array}
	\right)
	g(\theta_j-q_i \theta_{j-i}) d\theta_j
	= D_m Q_i
	\left(
	\begin{array}{rr}
	\cos m\theta_{j-i} & \sin m\theta_{j-i} \\
	\sin m\theta_{j-i} & -\cos m\theta_{j-i}
	\end{array}
	\right)
	Q_i,
	\end{align*}
	where $ D_m $ and $Q_i$ are the matrices defined in Lemma 1.
\end{lem}
\textbf{Proof}
Similar to the proof of Lemma 1, we have
\begin{align*}
&\int_{\Pi} 
\left(
\begin{array}{rr}
\cos m\theta_j & \sin m\theta_j\\
\sin m\theta_j & -\cos m\theta_j
\end{array}
\right)
g(\theta_j-q_i \theta_{j-i}) d\theta_j \\
=&
\int_{\Pi} 
\left(
\begin{array}{rr}
\cos m(s+q_i \theta_{j-i}) & \sin m(s+q_i \theta_{j-i})  \\
\sin m(s+q_i \theta_{j-i}) & -\cos m(s+q_i \theta_{j-i})
\end{array}
\right)
g(s) ds \\
=&
\int_{\Pi} 
\left(
\begin{array}{rr}
\cos ms & -\sin ms \\
\sin ms & \cos ms
\end{array}
\right)
g(s) ds 
\left(
\begin{array}{rr}
\cos q_i m\theta_{j-i} & \sin q_i m\theta_{j-i}  \\
\sin q_i m\theta_{j-i}  & -\cos q_i m\theta_{j-i}  
\end{array}
\right) \\
=
& 
D_m Q_i
\left(
\begin{array}{rr}
\cos m \theta_{j-i} & \sin m \theta_{j-i}  \\
\sin m \theta_{j-i} & -\cos m \theta_{j-i}
\end{array}
\right)
Q_i.
\end{align*}
\begin{flushleft}
\hfill $\square$
\end{flushleft}
~
\\
\medskip
 \noindent
 \textbf{Proof of Theorem~\ref{th:2}} See, for example, \citep[Theorem 2.7.1]{wei1990time} for the results of difference equations. \hfill $\square$
 \\
~
\medskip
\noindent
\textbf{Proof of Theorem~\ref{th:4}} 
Recall Assumption \ref{ass:mu=0} ensures the diagonal covariance matrices $\Gamma_j=\Gamma_{-j}=\mathrm{diag}(\gamma_{j,11},\gamma_{j,22})$. Then, the linear systems (\ref{forward}) and (\ref{backward}) become identical:
\begin{align}
\label{eq:linear systems}
\begin{bmatrix}
\Gamma_0 & \Gamma_{1} & \cdots & \Gamma_{s-1} \\
\Gamma_1 & \Gamma_{0} & \cdots & \Gamma_{s-2} \\
\vdots & \vdots & \ddots
& \vdots \\
\Gamma_{s-1} & \Gamma_{s-2} & \cdots & \Gamma_0    
\end{bmatrix}
\begin{bmatrix}
  \Xi_{1} \\
  \Xi_{2} \\
  \vdots \\
  \Xi_{s}
\end{bmatrix}
=
\begin{bmatrix}
\Gamma_{1} \\
\Gamma_{2} \\
\vdots \\
\Gamma_{s}
\end{bmatrix}. 
\end{align}
Note that the all roots of (\ref{eq:linear systems}) are diagonal matrices, which can be written as $\Xi_j = \mathrm{diag}(\xi_{j,11},\xi_{j,22})$. Then, the linear system (\ref{eq:linear systems}) decomposes into two systems:  
\begin{align*}
\begin{bmatrix}
\gamma_{0,11} & \gamma_{1,11} & \cdots & \gamma_{s-1,11} \\
\gamma_{1,11} & \gamma_{0,11} & \cdots & \gamma_{s-2,11} \\
\vdots & \vdots & \ddots
& \vdots \\
\gamma_{s-1,11} & \gamma_{s-2,11} & \cdots & \gamma_{0,11}  
\end{bmatrix}
\begin{bmatrix}
  \xi_{1,11} \\
  \xi_{2,11} \\
  \vdots \\
  \xi_{s,11}
\end{bmatrix}
=
\begin{bmatrix}
\gamma_{1,11} \\
\gamma_{2,11} \\
\vdots \\
\gamma_{s,11}
\end{bmatrix}
\end{align*}
and
\begin{align*}
\begin{bmatrix}
\gamma_{0,22} & \gamma_{1,22} & \cdots & \gamma_{s-1,22} \\
\gamma_{1,22} & \gamma_{0,22} & \cdots & \gamma_{s-2,22} \\
\vdots & \vdots & \ddots
& \vdots \\
\gamma_{s-1,22} & \gamma_{s-2,22} & \cdots & \gamma_{0,22}  
\end{bmatrix}
\begin{bmatrix}
  \xi_{1,22} \\
  \xi_{2,22} \\
  \vdots \\
  \xi_{s,22}
\end{bmatrix}
=
\begin{bmatrix}
\gamma_{1,22} \\
\gamma_{2,22} \\
\vdots \\
\gamma_{s,22}
\end{bmatrix}.
\end{align*}
Now, under Assumption \ref{ass:mu=0}, (\ref{CPACF}) becomes 
\begin{align*}
\psi_s^{(C)}=  \frac{ (\gamma_{s,11}-\xi_{1,11}\gamma_{s-1,11}-\ldots-\xi_{s-1,11}\gamma_{1,11})}
{(\gamma_{0,11}-\xi_{1,11}\gamma_{1,11}-\ldots-\xi_{s-1,11}\gamma_{s-1,11})} 
\times 
\frac{(\gamma_{s,22}-\xi_{1,22}\gamma_{s-1,22}-\ldots-\xi_{s-1,22}\gamma_{1,22})}
{(\gamma_{0,22}-\xi_{1,22}\gamma_{1,22}-\ldots-\xi_{s-1,22}\gamma_{s-1,22})}.
\end{align*}
Similar to the result of \citep[Section 2.3]{wei1990time}, it reduces to 
\begin{align*}
\psi_s^{(C)}
=&
\frac{
\begin{vmatrix}
 \gamma_{0,11} & \gamma_{1,11}& \cdots & \gamma_{s-2,11} & \gamma_{1,11} \\
  \gamma_{1,11} & \gamma_{0,11} & \cdots &  \gamma_{s-3,11} &  \gamma_{2,11} \\
  \vdots & \vdots & & \vdots & \vdots \\ 
  \gamma_{s-1,11} & \gamma_{s-2,11} & \cdots &  \gamma_{1,11} &  \gamma_{s,11} 
 \end{vmatrix}}
 {
 \begin{vmatrix}
 \gamma_{0,11} & \gamma_{1,11}& \cdots & \gamma_{s-2,11} & \gamma_{s-1,11} \\
  \gamma_{1,11} & \gamma_{0,11} & \cdots &  \gamma_{s-3,11} &  \gamma_{s-2,11} \\
  \vdots & \vdots & & \vdots & \vdots \\ 
  \gamma_{s-1,11} & \gamma_{s-2,11} & \cdots &  \gamma_{1,11} &  \gamma_{0,11} 
 \end{vmatrix}
 } 
 \frac{
 \begin{vmatrix}
 \gamma_{0,22} & \gamma_{1,22}& \cdots & \gamma_{s-2,22} & \gamma_{1,22} \\
  \gamma_{1,22} & \gamma_{0,22} & \cdots &  \gamma_{s-3,22} &  \gamma_{2,22} \\
  \vdots & \vdots & & \vdots & \vdots \\ 
  \gamma_{s-1,22} & \gamma_{s-2,22} & \cdots &  \gamma_{1,22} &  \gamma_{s,22} 
 \end{vmatrix}}
 {
 \begin{vmatrix}
 \gamma_{0,22} & \gamma_{1,22}& \cdots & \gamma_{s-2,22} & \gamma_{s-1,22} \\
  \gamma_{1,22} & \gamma_{0,22} & \cdots &  \gamma_{s-3,22} &  \gamma_{s-2,22} \\
  \vdots & \vdots & & \vdots & \vdots \\ 
  \gamma_{s-1,22} & \gamma_{s-2,22} & \cdots &  \gamma_{1,22} &  \gamma_{0,22} 
 \end{vmatrix}},
\end{align*}
which is identical to (\ref{eq:CPACF(det form)}). 
\hfill $\square$
\\~%
 \\~%
\medskip
\noindent
\textbf{Proof of Theorem~\ref{th:3}} Let $X_{1,t}$ and $X_{2,t}$ be two stationary AR($p$) processes, defined as
\begin{align*}
\phi_1(B) X_{1,t} = \tilde{u}_{1,t}~~~
\textrm{and}~~~
\phi_2(B) X_{2,t} = \tilde{u}_{2,t},
\end{align*}
where $  \tilde{u}_{1,t}$ and $\tilde{u}_{2,t}$ are uncorrelated processes with zero means, and some constant variances $\sigma_{\tilde{u}_1}^2$ and $\sigma_{\tilde{u}_2}^2$, respectively. 
Then, the autocovariances of $X_{1,t}$ and $X_{2,t}$ also satisfies (\ref{eq:arp}), which is representing the autocovariance relationships of $\cos\Theta_t$ and $\sin\Theta_t$. 
From the second-order stationarity of the processes $X_{1,t}$ and $X_{2,t}$, we have
\begin{align*}
\textrm{var} (X_{1,t}) = 
\frac{ \sigma_{\tilde{u}_{1}}^2}{ 1-2 \rho_1\sum_{i=1}^{p} a_i \gamma_{i,11}}
\end{align*}
and
\begin{align*}
\textrm{var} (X_{2,t}) = 
\frac{ \sigma_{\tilde{u}_{2}}^2}{ 1-2 \rho_1\sum_{i=1}^{p} q_ia_i \gamma_{i,22}}.
\end{align*}
As the circular uniform marginal assumption of $f$, we have $\textrm{var} (\cos\Theta_{t})=\textrm{var} (\sin\Theta_{t})=1/2$. Therefore, if we set $\textrm{var} (X_{1,t})=\textrm{var} (X_{2,t})=1/2$, that is, if we set 
\begin{align*}
\sigma_{\tilde{u}_{1}}^2
= \frac{1-2 \rho_1\sum_{i=1}^{p} a_i \gamma_{i,11}}{2} 
~~
\textrm{and}
~~
 \sigma_{\tilde{u}_{2}}^2 = 
  \frac{1-2 \rho_1\sum_{i=1}^{p} q_ia_i \gamma_{i,22}}{2}, 
\end{align*}
the processes $X_{1,t}$ and $X_{2,t}$ have the completely same autocovariance structures as those of the processes $\cos\Theta_t$ and $\sin\Theta_t$, respectively.  
Then, the spectral densities for the process $\cos\Theta_t$ and $\sin\Theta_t$ are the same as those of the processes $X_{1,t}$ and $X_{2,t}$, which are 
\begin{align*}
f_{X_1}(\omega) =
 \frac{1-2 \rho_1\sum_{i=1}^{p} a_i \gamma_{i,11}}{4\pi  | \phi_1(e^{-\mathrm{i}\omega})|^2}
\end{align*}
and
\begin{align*}
f_{X_2}(\omega) =
 \frac{1-2 \rho_1\sum_{i=1}^{p} q_ia_i \gamma_{i,22}}{4\pi  | \phi_2(e^{-\mathrm{i}\omega})|^2}.
\end{align*}
Using the convolution theorem of Fourier transforms, the $k$-th circular autocovariance function becomes
\begin{align*}
\gamma_{k, 11}\gamma_{k, 22} 
= &
\left(
\int_{-\pi}^{\pi} e^{\mathrm{i}k\omega_1} f_{X_1}(\omega_1) d \omega_1
\right)
\left(
 \int_{-\pi}^{\pi} e^{\mathrm{i}k\omega_2} f_{X_2}(\omega_2) d \omega_2
\right) \\
=&
\int_{-\pi}^{\pi}
\left(
\int_{-\pi}^{\pi}
 f_{X_1}(\omega_1)  
 f_{X_2}(\omega_2- \omega_1) d\omega_1 \right)
 e^{\mathrm{i} k \omega_2} d\omega_2. 
\end{align*}
This implies that the spectral density of $\Theta_t$ is expressed as that in the
theorem.\hfill $\square$
\\
~%
\\
\noindent
\textbf{Proof of Theorem \ref{th5}}
As the first- and second-order stationarity results stated by Theorems 1 and 2 together with the definition of the Markov kernel density (\ref{eq:MTD}) and its autocovariance function is absolutely summable such that $\sum_{k=-\infty}^{\infty} |\det \Gamma_k| < \infty$, the process (\ref{eq:MTD}) is strictly stationary and ergodic. The strongly consistent result follows from \citep[Theorem 8.7]{douc2014nonlinear}, as Assumption 2(b) and (c) imply that the Markov kernel density defined by (\ref{eq:MTD}) is continuous and its logarithm is bounded above by 
$\sup_{\bm{\eta}\in\bm{H}}E_{\eta_0}\{\ln (\sum_{i=1}^p a_{i} g(\Theta_t -q_{i0} \Theta_{t-i}; \rho))\} < \infty$. 
\hfill $\square$
\\
\\
~%
\noindent
\textbf{Proof of Theorem \ref{th6}}
From Assumption 3(b), the continuity of $g(\cdot)$ implies $\sum_{i=1}^p a_{i} g(\theta_t -q_{i0} \theta_{t-i}; \rho)$ is continuous, which yields following
\begin{align*}
\left.
\frac{\partial}{\partial\bm{\eta}} \ell_n(\bm{\eta}, \bm{q}_0)
\right|_{\hat{\bm{\eta}}_n} =\bm{0}.
\end{align*}
In addition, we observe the following expansion as
\begin{align*}
\left.
\frac{\partial}{\partial\bm{\eta}} \ell_n(\bm{\eta}, \bm{q}_0)
\right|_{\bm{\eta}_0} 
+
\left.
\frac{\partial^2}{\partial\bm{\eta}\partial\bm{\eta}^T} \ell_n(\bm{\eta}, \bm{q}_0) 
\right|_{\bm{\eta}_*} 
(\hat{\bm{\eta}}_n -\bm{\eta}_0)
=\bm{0},
\end{align*}
where $\bm{\eta}_*$ satisfies $\Vert \bm{\eta}_*- \bm{\eta}_0\Vert \leq \Vert \hat{\bm{\eta}}_n- \bm{\eta}_0\Vert $. Then we observe
\begin{align*}
\sqrt{n} (\hat{\bm{\eta}}_n -\bm{\eta}_0) = 
\left[
-
\left.
\frac{1}{n}\frac{\partial^2}{\partial\bm{\eta}\partial\bm{\eta}^T} \ell_n(\bm{\eta}, \bm{q}_0) 
\right|_{\bm{\eta}_*} 
\right]^{-1}
\left.
\frac{1}{\sqrt{n}}
\frac{\partial}{\partial\bm{\eta}} \ell_n(\bm{\eta}, \bm{q}_0)
\right|_{\bm{\eta}_0} .
\end{align*}
Using a law of large numbers, we have
\begin{align*}
-\frac{1}{n}\frac{\partial^2}{\partial\bm{\eta}\partial\bm{\eta}^T} \ell_n(\bm{\eta}_*, \bm{q}_0) 
\to_p I(\bm{\eta}_0)
\end{align*} 
where $I(\bm{\eta}_0)$ is defined by Assumption 3(d).
A central limit theorem implies
\begin{align*}
\frac{1}{\sqrt{n}}
\frac{\partial}{\partial\bm{\eta}} \ell_n(\bm{\eta}, \bm{q}_0)
\mathop{\to}_{d}
N\left( \bm{0},
\bm{J}(\bm{\eta}_0) \right),
\end{align*}
see, for details, \citep[Theorem~8.13]{douc2014nonlinear}.
This, together with Slutsky's lemma, completes the proof of the theorem.
\hfill $\square$

\end{document}